\documentclass[]{article}
\usepackage{proceed2e}
\usepackage{graphicx}
\usepackage{subfigure}
\usepackage{algorithm}
\usepackage{algorithmic}
\usepackage{amsmath}
\usepackage{amssymb}
\usepackage{xspace}
\usepackage{color}

\title{Modeling Social Networks with Node Attributes using the\\ Multiplicative Attribute Graph Model}

\newcommand{\hide}[1]{}

\newcommand{\rev}[1]{{#1}}
\newcommand{\eg}{\emph{e.g.}\xspace}
\newcommand{\ie}{\emph{i.e.}\xspace}
\newcommand{\xhdr}[1]{{\bf {#1}.}}
\newcommand{\MAG}{MAG model\xspace}
\newcommand{\MMSB}{MMSB}
\newcommand{\MAGFIT}{\textsc{MagFit}\xspace}
\newtheorem{problem}{Problem}
\newcommand{\denselist}{\itemsep-3pt \topsep-5pt \partopsep-5pt}

\author{ {\bf Myunghwan Kim} \\
Stanford University\\
Stanford, CA 94305 \\
\And
{\bf Jure Leskovec}  \\
Stanford University  \\
Stanford, CA 94305 \\
}

\begin{document}

\maketitle
\begin{abstract}
Networks arising from social, technological and natural domains exhibit rich
connectivity patterns and nodes in such networks are often labeled with
attributes or features. We address the question of modeling the structure of
networks where nodes have attribute information.
%
%
We present a Multiplicative Attribute Graph (MAG) model that considers nodes
with categorical attributes and models the probability of an edge as the
product of individual attribute link formation affinities. We develop a
scalable variational expectation maximization parameter estimation method.
Experiments show that MAG model reliably captures network connectivity as well
as provides insights into how different attributes shape the network structure.


\end{abstract}

\section{Introduction}
\label{sec:intro}
\rev{Social and biological systems can be modeled as interaction networks}
\rev{where nodes and edges represent entities and interactions.}
Viewing real systems as networks led to discovery of
\rev{underlying organizational principles~\cite{faloutsos99powerlaw,watts98smallworld}}
as well as to high impact applications~\cite{page98pagerank}.
As organizational principles of networks are discovered,
\rev{questions are as follow:}
\rev{Why are networks organized the way they are?}
How can we model this?

Network modeling has rich history and can be roughly divided into two streams.
First are the explanatory ``mechanistic''
models~\cite{kumar00stochastic,jure05dpl} that posit simple generative
mechanisms that lead to networks with realistic connectivity
patterns.
\rev{For example,}
the Copying model~\cite{kumar00stochastic} states a
simple rule where a new node joins the network, randomly picks an existing node
and links to some of its neighbors. One can {\em prove} that under this generative
mechanism networks with power-law degree distributions naturally emerge.
Second line of work are statistical models of network
structure~\cite{airoldi07blockmodel,hoff02latent,robins07egrm,sarkar05dynamic}
which are usually accompanied by model parameter estimation procedures and have
proven to be useful for hypothesis testing. However, such models are often
analytically untractable as they do not lend themselves to mathematical
analysis of structural properties of networks that emerge from the models.

Recently a new line of work~\cite{Palla2010,young07dotprod} has emerged.
It develops network models that are analytically tractable in a sense that one can
mathematically analyze structural properties of networks that emerge from the
models as well as statistically meaningful in a sense that there exist efficient
parameter estimation techniques.
\rev{For instance, Kronecker graphs model~\cite{jure05kronecker} can be mathematically proved that it gives rise to networks with a small diameter, giant connected component, and so on~\cite{mahdian07kronecker,jure10kronecker}. Also, it can be fitted to real networks~\cite{jure07kronfit} to reliably mimic their structure.}

However, the above models focus only on modeling the network structure while
not considering information about properties of the nodes of the network. Often
nodes have features or attributes associated with them. And the question is how
to characterize and model the interactions between the node properties and
the network structure.
\rev{For instance,}
users in a online social network have
profile information
\rev{like age and gender,}
and we are interested in
modeling how these attributes interact to give rise to the observed network
structure.

We present the {\em Multiplicative Attribute Graphs (MAG)} model that naturally
captures interactions between the node attributes and the observed network
structure. The model considers nodes with categorical attributes and the
probability of an edge between a pair of nodes depends on the individual
attribute link formation affinities.
The MAG model is analytically tractable in a sense that we can prove that networks
arising from the model exhibit connectivity patterns that are also found in real-world
networks~\cite{mh10mag}. For example, networks arising from the model have
heavy-tailed degree distributions, small diameter and unique giant connected
component~\cite{mh10mag}. Moreover, the MAG model captures homophily (\ie, tendency to link
to similar others) as well as heterophily (\ie, tendency to link to
different others) of different node attributes.

In this paper we develop \MAGFIT, a scalable parameter estimation method for the MAG
model. We start by defining the generative interpretation of the model and then
cast the model parameter estimation as a maximum likelihood problem. Our approach
is based on the variational expectation maximization framework and nicely
scales to large networks.
Experiments on several real-world networks demonstrate that the MAG model
reliably captures the network connectivity patterns and outperforms present
state-of-the-art methods. Moreover, the model parameters have natural
interpretation and provide additional insights into
\rev{how node attributes shape the structure of networks.}

\section{Multiplicative Attribute Graphs}
\label{sec:problem}
\newcommand{\A}{\mathbb{A}}
\newcommand{\AL}{F_{\{\cdot l\}}}
\newcommand{\ALP}{F_{\{\cdot l'\}}}
\newcommand{\AI}[1]{F_{{#1}}}
\newcommand{\Y}{\mathbb{A}}
\newcommand{\BERLIK}[2]{{#1}^{{#2}} \left(1 - {#1}\right)^{1 - {#2}}}
\newcommand{\BETALIK}[3]{\frac{\Gamma({#2})\Gamma({#3})}{\Gamma({#2}+{#3})}{#1}^{{#2}-1} \left(1 - {#1}\right)^{{#3} - 1}}
\newcommand{\LL}{\mathcal{L}}
\newcommand{\EQ}[1]{Eq.~(\ref{#1})}
\newcommand{\LIKJOIN}{P\left(A, F | \mu, \Theta\right)}
\newcommand{\LIKALL}{P(A| \mu, \Theta)}

The Multiplicative Attribute Graphs model (MAG)~\cite{mh10mag} is a class of
generative models for networks with node attributes. MAG combines categorical
node attributes with their affinities to compute\hide{estimate} the probability of a link.
For example, some node attributes (\eg, political affiliation) may have
positive affinities in a sense that same political view increases probability
of being linked (\ie, homophily), while other attributes may have negative
affinities, \ie, people are more likely to link to others with a different
value of that attribute.

\begin{figure}
\centering
\includegraphics[width=0.5\textwidth]{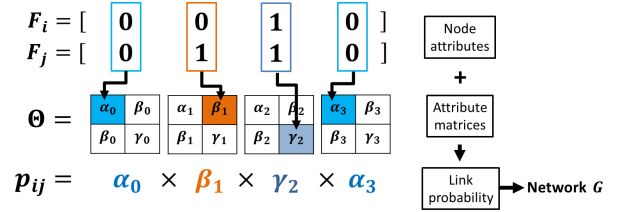}
\vspace{-5mm}
\caption{{\em Multiplicative Attribute Graph (MAG) model}. Each node $i$ has
categorical attribute vector $F_i$. The probability $p_{ij}$ of edge $(i,j)$
is then determined by attributes ``selecting'' appropriate the entries of
attribute affinity matrices $\Theta_l$. }
\vspace{-3mm}
\label{fig:mag}
\end{figure}

Formally, we consider a directed graph $A$
(represented by its binary adjacency matrix) on $N$ nodes. 
Each node $i$ has $L$ categorical attributes, $F_{i1}, \cdots, F_{iL}$ and each
attribute $l$ ($l = 1, \cdots, L$) is associated with affinity matrix
$\Theta_{l}$ which quantifies the affinity of the attribute to form a link .
Each entry $\Theta_{l}[k, k'] \in (0,1)$ of the affinity matrix indicates the
potential for a pair of nodes to form a link, given
the $l$-th attribute value $k$ of the first node and value $k'$ of the second node.
%
For a given pair of nodes, their attribute values ``select''
\rev{proper}
entries of affinity matrices, \ie, the first node's attribute selects a
``row'' while the second node's attribute value selects a ``column''. The link
probability is then defined as the product of the selected entries of affinity
matrices. Each edge $(i,j)$ is then included in the graph $A$ independently
with probability $p_{ij}$:
\begin{equation}
  p_{ij} := P(A_{ij} = 1) = \prod_{l = 1}^{L} \Theta_{l}[F_{il}, F_{jl}] \,.
  \label{eq:mag}
\end{equation}
%
%
Figure~\ref{fig:mag} illustrates the model. Nodes $i$ and $j$ have the binary
attribute vectors $[0, 0, 1, 0]$ and $[0, 1, 1, 0]$, respectively. We then select
the entries of the attribute matrices, $\Theta_{1}[0, 0]$, $\Theta_{2}[0, 1]$,
$\Theta_{3}[1, 1]$, and $\Theta_{4}[0, 0]$ and compute the link
probability $p_{ij}$ of link $(i, j)$ as a product of these selected entries.

Kim \& Leskovec~\cite{mh10mag} proved that the \MAG captures connectivity
patterns observed in real-world networks,
\rev{such as}
heavy-tailed (power-law or
log-normal) degree distributions, small diameters, unique giant connected
component and local clustering of the edges. They provided both
analytical and empirical evidence demonstrating that the \MAG effectively
captures the structure of real-world networks.

The \MAG~can handle attributes of any cardinality, however, for simplicity we limit
our discussion to binary attributes. Thus, every $F_{il}$ takes value of either $0$ or $1$,
and every $\Theta_{l}$ is a $2 \times 2$ matrix.

\xhdr{Model parameter estimation}
So far we have seen how given the node attributes $F$ and the corresponding
attribute affinity matrices $\Theta$ we generate a MAG network. Now we focus on
the reverse problem: Given a network $A$ and the number of attributes $L$ we
aim to estimate affinity matrices $\Theta$ and
node attributes $F$.

In other words, we aim to represent the given real network $A$ in the form of the \MAG
parameters: node attributes $F = \{F_{il}; i = 1, \cdots, N, ~l=1, \cdots, L\}$
and attribute affinity matrices $\Theta = \{\Theta_{l}; l = 1, \cdots, L\}$.
MAG yields a probabilistic adjacency matrix that independently assigns the link probability to
every pair of nodes, the likelihood $P(A | F, \Theta)$ of a given graph
(adjacency matrix) $A$ is the product of the edge probabilities over the
edges and non-edges of the network:
\begin{equation}
  P(A | F, \Theta) = \prod_{A_{ij} = 1} p_{ij} \prod_{A_{ij} = 0} (1 - p_{ij})
\end{equation}
and $p_{ij}$ is defined in \EQ{eq:mag}.

Now we can use the maximum likelihood estimation to find node attributes $F$
and their affinity matrices $\Theta$. Hence, ideally we would like to solve
%
\begin{equation}
  \arg \max_{F, \Theta} P(A | F, \Theta) \label{eq:llbasic} \,.
\end{equation}

However, there are several challenges with this problem formulation. First,
notice that \EQ{eq:llbasic} is a combinatorial problem of $O(LN)$ categorical
variables even when the affinity matrices $\Theta$ are fixed. Finding both $F$
and $\Theta$ simultaneously is
even harder.
Second, even if we could
solve this combinatorial problem, the model has
\rev{a lot of}
parameters
\rev{which may cause high variance.}

To resolve these challenges, we consider a simple generative model for
the node attributes.
%
%
We assume that the $l$-th attribute of each node is drawn 
\rev{from an {\em i.i.d.}}
Bernoulli distribution parameterized by $\mu_{l}$. This means that the $l$-th
attribute of every node takes value 1 with probability $\mu_{l}$, \ie, $F_{il} \sim
Bernoulli \left( \mu_{l} \right)$.

\begin{figure}
\centering
  \includegraphics[width=0.3\textwidth]{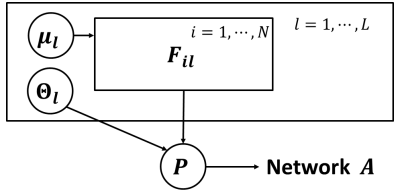}
  \caption{MAG model: Node attributes $F_{il}$ are sampled
  from $\mu_{l}$ and combined with
  affinity matrices $\Theta_{l}$ to generate a probabilistic adjacency
  matrix $P$.}
\label{fig:mag-plate}
\end{figure}

Figure~\ref{fig:mag-plate} illustrates the model in plate notation. First, node
attributes $F_{il}$ are generated by the corresponding Bernoulli distributions
$\mu_{l}$. By combining these node attributes with the affinity matrices
$\Theta_{l}$, the probabilistic adjacency matrix $P$ is formed. Network $A$ is then
generated by a series of coin flips where each edge $A_{ij}$ appears with
probability $P_{ij}$.

Even this simplified model provably generates networks with power-law
degree distributions, small diameter, and unique giant component~\cite{mh10mag}. The simplified model requires only $5L$ parameters (4 per each $\Theta_l$, 1 per $\mu_l$). Note that the number of
attributes $L$ can be thought of as constant or slowly increasing in the
number of nodes $N$ (\eg, $L=O(\log N)$)~\cite{bonato10waw,mh10mag}.



%
The generative model for node attributes slightly modifies the objective
function in \EQ{eq:llbasic}. We maintain the maximum likelihood approach, but
instead of directly finding attributes $F$ we now estimate parameters $\mu_l$
that then generate latent node attributes $F$.

We denote the log-likelihood $\log P(A | \mu, \Theta)$ as $\LL(\mu, \Theta)$
and aim to find $\mu = \{\mu_{l}\}$ and $\Theta = \{\Theta_{l}\}$ by
\rev{maximizing}
\begin{equation}
  \rev{\LL(\mu, \Theta) = \log \LIKALL = \log \sum_{F} \LIKJOIN \,.}
  \nonumber
\end{equation}
Note that since $\mu$ and $\Theta$ are linked through $F$ we have to sum over
all possible instantiations of node attributes $F$. Since $F$ consists of
$L\cdot N$ binary variables, the number of all possible instantiations of $F$
is $O(2^{LN})$, which makes computing $\LL(\mu, \Theta)$ directly intractable.
In the next section we will show how to quickly (but approximately) compute the summation.

To compute likelihood $\LIKJOIN$, we have to consider the likelihood of node
attributes. Note that each edge $A_{ij}$ is independent given the attributes
$F$ and each attribute $F_{il}$ is independent given the parameters $\mu_{l}$.
By this conditional independence and the fact that both $A_{ij}$ and $F_{il}$
follow Bernoulli distributions with parameters $p_{ij}$ and $\mu_{l}$ we
obtain
\begin{align}
  & P(A, F | \mu, \Theta) = P(A | F, \mu, \Theta) P(F | \mu, \Theta) \nonumber \\
  & = P(A | F, \Theta) P(F | \mu) \nonumber \\
  & = \prod_{A_{ij} = 1} p_{ij} \prod_{A_{ij} = 0}  (1-p_{ij})
  \prod_{F_{il} = 0} \mu_{l} \prod_{F_{il} = 1} (1 - \mu_{l})
  \label{eq:jointlik}
\end{align}
where $p_{ij}$ is defined in \EQ{eq:mag}.

\section{MAG Parameter Estimation}
\label{sec:algorithm}
\newcommand{\LIKPOS}{P(F | A, \mu, \Theta)}
\newcommand{\ALIKJOIN}{Q(F)}
\newcommand{\ATTRPOSF}[1]{Q_{\phi_{{#1}}} (F_{{#1}} | \phi_{{#1}})}
\newcommand{\MUPOSF}[1]{Q_{\gamma_{{#1}}} (\mu_{{#1}} | \gamma_{{#1}})}
\newcommand{\ATTRPOSA}[1]{Q_{\phi_{{#1}}}(F_{{#1}})}
\newcommand{\MUPOSA}[1]{Q_{\gamma_{{#1}}}}
\newcommand{\EXATTR}[2]{\mathbf{E}_{Q_{#1}}\left[ {#2} \right]}
\newcommand{\LOGBER}[2]{{#1}\log {#2} + (1 - {#1}) \log (1- {#2})}
\newcommand{\LOGA}{\log P(A | F, \Theta)}
\newcommand{\LOGEXA}[1]{\EXATTR{#1}{\LOGA}}
\newcommand{\LOGG}[1]{\tilde{P}_{#1}}
\newcommand{\WEIGHTC}[2]{w_{{#1}}^{(#2)}}
\newcommand{\WEIGHTR}[1]{\overrightarrow{w_{{#1}}}}
\newcommand{\WEIGHTL}[1]{\overleftarrow{w_{{#1}}}}
\newcommand{\PFUNC}[1]{\tilde{P_{il}}\left({#1}\right)}
\newcommand{\LQ}{\mathcal{L}_{Q}}
\newcommand{\LMU}{\mathcal{L}_{\mu_{l}}}
\newcommand{\LTH}{\mathcal{L}_{\Theta}}
\newcommand{\PLQ}[1]{\frac{\partial \mathcal{L}_{Q}}{\partial {#1}}}
\newcommand{\MI}{\mbox{MI}}
\newcommand{\PMI}[1]{\frac{\partial \mbox{MI}}{\partial {#1}}}


Now, given a network $A$, we aim to estimate the parameters $\mu_l$ of the node
attribute model as well as the attribute affinity matrices $\Theta_l$. We
regard the actual node attribute values $F$ as latent variables and use the
expectation maximization framework.

We present the approximate method to solve the problem by developing a
variational Expectation-Maximization (EM) algorithm. We first derive the lower
bound $\LL_Q(\mu, \Theta)$ on the true log-likelihood $\LL(\mu, \Theta)$ by
introducing the variational distribution $Q(F)$ parameterized by variational
parameters $\phi$.
\rev{Then, we}
indirectly maximize $\LL(\mu, \Theta)$ by maximizing
its lower bound $\LL_Q(\mu, \Theta)$.
%
In the E-step, we estimate $Q(F)$ by maximizing
\rev{$\LL_Q(\mu, \Theta)$}
over the variational parameters $\phi$.
In the M-step, we maximize the lower bound $\LL_Q(\mu, \Theta)$
over the \MAG parameters ($\mu$ and $\Theta$) to approximately maximize the actual log-likelihood $\LL(\mu,
\Theta)$. We alternate between E- and M-steps until the parameters converge.

%
%
%
%
%

\xhdr{Variational EM}
Next we introduce the distribution $Q(F)$ parameterized by
variational parameters $\phi$. The idea is to define an easy-to-compute $Q(F)$
that allows us to compute the lower-bound $\LL_Q(\mu, \Theta)$ of the true
log-likelihood $\LL(\mu, \Theta)$. Then instead of maximizing the hard-to-compute $\LL$, we  maximize $\LL_Q$.


We now show that in order to make the gap between the lower-bound $\LL_Q$ and the original log likelihood $\LL$ small we should find the easy-to-compute $Q(F)$ that closely approximates $P(F | A, \mu, \Theta)$. For now we keep $Q(F)$ abstract and precisely define it later.
%
%

We begin by computing the lower bound $\LL_Q$ in terms of
$Q(F)$. We plug $Q(F)$ into $\LL(\mu, \Theta)$ as follows:

\begin{small}
\begin{align}
\LL(\mu, \Theta) & = \log \sum_{F} P(A, F | \mu, \Theta) \nonumber \\
& = \log \sum_{F} Q(F) \frac{P(A, F | \mu, \Theta)}{Q(F)} \nonumber \\
& = \log \EXATTR{}{ \frac{P(A, F | \mu, \Theta)}{Q(F)} } \,.
\label{eq:expll}
\end{align}
\end{small}
\vspace{-5mm}

As $\log x$ is a concave function, by Jensen's inequality,
\begin{equation}
  \log \EXATTR{}{ \frac{P(A, F | \mu, \Theta)}{Q(F)} }
  \geq \EXATTR{}{\log \frac{P(A, F | \mu, \Theta)}{Q(F)}} \,.\nonumber
\end{equation}
Therefore, by taking
\begin{equation}
  \LQ(\mu, \Theta) = \EXATTR{}{\log P(A, F | \mu, \Theta) - \log Q(F)} \,,
  \label{eq:lqdef}
\end{equation}
$\LQ(\mu, \Theta)$ becomes the lower bound on $\LL(\mu, \Theta)$.


Now the question is how to set $Q(F)$ so that we make the gap between $\LQ$ and $\LL$ as small as possible.
The lower bound $\LQ$ is tight when the proposal distribution $Q(F)$
becomes close to the true posterior distribution $P(F | A, \mu, \Theta)$ in the
KL divergence. More precisely, since $P(A | \mu, \Theta)$ is
independent of $F$, $\LL(\mu, \Theta) = \log P(A | \mu,
\Theta) = \EXATTR{}{\log P(A | \mu, \Theta)}$. Thus, the gap between
$\LL$ and $\LQ$ is
\begin{align}
  & \LL(\mu, \Theta) - \LQ(\mu, \Theta) \nonumber \\
  & = \log P(A | \mu, \Theta) - \EXATTR{}{\log P(A, F | \mu, \Theta) - \log Q(F)} \nonumber \\
  & = \EXATTR{}{\log P(A | \mu, \Theta) - \log P(A, F | \mu, \Theta) + \log Q(F)} \nonumber \\
  & = \EXATTR{}{\log P(F | A, \mu, \Theta) - \log Q(F)} \, ,
  \nonumber
\end{align}
which means that the gap between $\LL$ and $\LQ$ is exactly the KL divergence between the proposal distribution $Q(F)$ and the true posterior distribution $P(F | A, \mu, \Theta)$.

%
%
%
%

Now we know how to choose $Q(F)$ to make the gap small. We want $Q(F)$ that is easy-to-compute and at the same time closely approximates $P(F | A, \mu, \Theta)$. We propose the following $Q(F)$ parameterized by $\phi$:
\begin{align}
  F_{il} & \sim Bernoulli(\phi_{il}) \nonumber \\
  Q_{il}(F_{il}) & = \BERLIK{\phi_{il}}{F_{il}} \nonumber \\
  Q(F) & = \prod_{i, l} Q_{il}(F_{il})
\label{eq:qdef}
\end{align}
where $\phi = \{\phi_{il}\}$ are variational parameters and $F = \{F_{il}\}$.
Our $Q(F)$ has several advantages. First, the computation of
$\LQ$ for fixed model parameters $\mu$ and $\Theta$ is tractable
because $\log P(A, F | \mu, \Theta) - \log Q(F)$ in \EQ{eq:lqdef} is
separable in terms of $F_{il}$.
This means that we are able to update each $\phi_{il}$ in turn to maximize
$\LQ$ by fixing all the other parameters: $\mu$, $\Theta$ and all $\phi$ except the
given $\phi_{il}$.
%
%
Furthermore, since each $\phi_{il}$ represents the approximate
posterior distribution of $F_{il}$ given the network, we can estimate each
attribute $F_{il}$ by $\phi_{il}$.

\xhdr{Regularization by mutual information}
In order to improve the robustness of MAG parameter estimation procedure, we enforce that each attribute is independent of others.
%
The maximum likelihood estimation cannot guarantee the independence between the
node attributes and so the solution might converge to local optima where
the attributes are correlated. To prevent this, we add a penalty term that aims to minimize the mutual information (\ie, maximize the entropy) between pairs of attributes.

Since the distribution for each attribute $F_{il}$ is defined by $\phi_{il}$,
we define the mutual information between a pair of attributes in terms of
$\phi$. We denote this mutual information as $\MI(F) = \sum_{l \neq l'}
\MI_{ll'}$ where $\MI_{ll'}$ represents the mutual information between the
attributes $l$ and $l'$. We then regularize the log-likelihood with the mutual
information term. We arrive to the following \MAGFIT optimization problem that
we actually solve
\begin{equation}
  \arg \max_{\phi, \mu, \Theta} \LQ(\mu, \Theta) - \lambda \sum_{l \neq l'} \MI_{ll'} \, .
  \label{eq:llmi}
\end{equation}
We can quickly compute the mutual information $\MI_{ll'}$ between attributes
$l$ and $l'$. Let $\AL$ denote a random variable representing the value of
attribute $l$\hide{vector of values of attribute $l$}. Then, the probability
$P(\AL = x)$ that attribute $l$ takes value $x$ is computed by averaging
$Q_{il}(x)$ over $i$. Similarly, the joint probability $P(\AL = x, \ALP = y)$
of attributes $l$ and $l'$ taking values $x$ and $y$
\rev{can be computed}
given $Q(F)$.
We compute $\MI_{ll'}$ using $Q_{il}$ defined in \EQ{eq:qdef} as
follows:
\begin{align}
  & p_{l}(x) := P(\AL = x) = \frac{1}{N}\sum_{i} Q_{il}(x) \nonumber\\
  & p_{ll'}(x, y) := P(\AL = x, \ALP = y) = \frac{1}{N}\sum_{i} Q_{il}(x) Q_{il'}(y) \nonumber  \\
  & \MI_{ll'} = \sum_{x,y \in \{0, 1\}} p_{ll'}(x, y)\log \left( \frac{p_{ll'}(x, y)}{p_{l}(x)p_{l'}(y)} \right) \,.
  \label{eq:midef}
\end{align}
%


\xhdr{The \MAGFIT algorithm}
To solve the regularized \MAGFIT problem in \EQ{eq:llmi}, we use the EM
algorithm which maximizes the lower bound $\LQ(\mu, \Theta)$ regularized by the
mutual information. In the E-step, we reduce the gap between the original
likelihood $\LL(\mu, \Theta)$ and its lower bound $\LQ(\mu, \Theta)$ as well as
minimize the mutual information between pairs of attributes. By fixing the
model parameters $\mu$ and $\Theta$, we update $\phi_{il}$ one by one using a
gradient-based method.
In the M-step, we then maximize $\LQ(\mu, \Theta)$ by updating the model
parameters $\mu$ and $\Theta$. We repeat E- and M-steps until all the
parameters $\phi$, $\mu$, and $\Theta$ converge.
Next we briefly overview the E- and the M-step. We give further details
in Appendix.

\xhdr{Variational E-Step}
In the E-step, we consider model parameters $\mu$ and $\Theta$
as given and we aim to find the values of variational parameters $\phi$ that
maximize $\LQ(\mu, \Theta)$ as well as minimize the mutual information
$\mbox{MI}(F)$.
%
We use the stochastic gradient method to update variational parameters
$\phi$. We randomly select a batch of entries in $\phi$ and update them by their gradient values of the objective function in  \EQ{eq:llmi}. We repeat this procedure until parameters $\phi$ converge.

First,
by computing $\PLQ{\phi_{il}}$ and $\frac{\partial \MI}{\partial\phi_{il}}$,
we obtain the gradient $\nabla_{\phi} \left( \LQ(\mu, \Theta) - \lambda \MI(F)
\right)$
(see Appendix for details).
%
%
Then we choose a batch of $\phi_{il}$ at random
and update them by $\PLQ{\phi_{il}} - \lambda \PMI{\phi_{il}}$ in each step.
The mutual information regularization term typically works in the opposite direction of the
likelihood. Intuitively, the regularization prevents the
solution from being stuck in the local optimum where the node attributes
are correlated.
Algorithm~\ref{alg:estep} gives the pseudocode.

\newcommand{\AF}[1]{\mbox{\textsc{#1}}}
\begin{algorithm}[t]
\small
\caption{\AF{MagFit-VarEStep}($A, \mu, \Theta$)}
\label{alg:estep}
\begin{algorithmic}
\STATE Initialize $\phi^{(0)}=\{\phi_{il}: i = 1, \cdots, N, \quad l = 1, \cdots, L \}$
\STATE ~~
\FOR{$t \gets 0$ to $T-1$}
	\STATE $\phi^{(t+1)} \gets \phi^{(t)}$
	\STATE Select $S \subset \phi^{(t)}$ with $|S| = B$
	\FOR{$\phi_{il}^{(t)} \in S$}
    \STATE Compute $\PLQ{\phi_{il}}$
		\STATE $\PMI{\phi_{il}} \gets 0$
		\FOR{$l' \neq l$}
			\STATE Compute $\frac{\partial \MI_{ll'}}{\partial \phi_{il}}$
			\STATE $\PMI{\phi_{il}} \gets \PMI{\phi_{il}} + \frac{\partial \MI_{ll'}}{\partial \phi_{il}}$
		\ENDFOR
		\STATE $\phi_{il}^{(t+1)} \gets \phi_{il}^{(t)} + \eta (\PLQ{\phi_{il}} - \lambda \PMI{\phi_{il}})$
	\ENDFOR
\ENDFOR
\end{algorithmic}
\end{algorithm}

\begin{algorithm}[t]
\small
\caption{\AF{MagFit-VarMStep}($\phi$, $G$, $\Theta^{(0)}$)}
\label{alg:mstep}
\begin{algorithmic}
\STATE ~~
\FOR{$l \gets 1$ to $L$}
	\STATE $\mu_{l} \gets \frac{1}{N} \sum_{i} \phi_{il}$
\ENDFOR
\STATE ~~
\FOR{$t \gets 0$ to $T-1$}
	\FOR{$l \gets 1$ to $L$}
		\STATE $\Theta^{(t+1)}_{l} \gets \Theta^{(t)}_{l} + \eta \nabla_{\Theta_{l}} \LQ$
	\ENDFOR
\ENDFOR
%
\end{algorithmic}
\end{algorithm}

\xhdr{Variational M-Step}
In the E-step, we introduced the variational distribution $Q(F)$ parameterized
by $\phi$ and approximated the posterior distribution $P(F| A, \mu, \Theta)$ by
maximizing $\LQ(\mu, \Theta)$ over $\phi$.
%
In the M-step, we now fix $Q(F)$,
\ie, fix the variational parameters $\phi$, and update the model parameters
$\mu$ and $\Theta$ to maximize $\LQ$.

%

First, in order to maximize $\LQ(\mu, \Theta)$ with respect to $\mu$, we need to maximize $\LMU = \sum_{i} \EXATTR{il}{\log P(F_{il} |
\mu_{l})}$ for each $\mu_{l}$.
By definitions in \EQ{eq:jointlik} and (\ref{eq:qdef}), we obtain
\[
  \LMU = \sum_{i} \left( \phi_{il} \mu_{il} + (1-\phi_{il})(1-\mu_{il}) \right) \,.
\]
Then $\LMU$ is maximized when
\[
  \frac{\partial \LMU}{\partial \mu_{l}} = \sum_{i} \phi_{il} - N = 0
\]
where $\mu_{l} = \frac{1}{N} \sum_{i} \phi_{il}$.

Second, to maximize $\LQ(\mu, \Theta)$ with respect to $\Theta_{l}$, we
maximize $\LTH = \EXATTR{}{\log P(A, F | \mu, \Theta) - \log Q(F)}$.
We first obtain the gradient
%
\begin{align}
  \nabla_{\Theta_{l}} \LTH = \sum_{i, j} \nabla_{\Theta_{l}} \EXATTR{i, j}{\log P(A_{ij} |F_{i}, F_{j}, \Theta)}
  \label{eq:mstep-theta}
\end{align}
and then use a gradient-based method to optimize $\LQ(\mu, \Theta)$ with regard to $\Theta_{l}$.
Algorithm~\ref{alg:mstep} gives details for optimizing $\LQ(\mu, \Theta)$ over $\mu$ and $\Theta$.

\xhdr{Speeding up \MAGFIT}
So far we described how to apply the variational EM algorithm to MAG model parameter estimation. However, both E-step and M-step are infeasible when the number of nodes $N$ is large. In particular, in the E-step, for each update of $\phi_{il}$, we have to
compute the expected log-likelihood value of every entry in the $i$-th row and
column of the adjacency matrix $A$.
It takes $O(LN)$ time to do this, so overall $O(L^2 N^2)$ time is needed to update all $\phi_{il}$. Similarly, in
the M-step, we need to sum up the gradient of $\Theta_{l}$ over every pair of
nodes (as in \EQ{eq:mstep-theta}). Therefore, the M-step requires $O(L
N^2)$ time and so it takes $O(L^2 N^2)$ to run a single iteration of EM. Quadratic dependency in the number of attributes $L$ and the number of nodes $N$ is infeasible for the size of the networks that we aim to work with here.

To tackle this, we make the following observation. Note that both
\EQ{eq:mstep-theta} and computation of $\PLQ{\phi_{il}}$ involve the sum of
\rev{expected values of the log-likelihood or the gradient.}
If we can quickly approximate this sum of the expectations,
we can dramatically reduce the computation time. As real-world networks are
sparse in a sense that most of the edges do not exist in the network, we can
break the summation into two parts --- a fixed part that ``pretends'' that the
network has no edges and the adjustment part \rev{that} takes into account the edges
that actually exist in the network.

For example, in the M-step we can separate \EQ{eq:mstep-theta} into two parts,
the first term\hide{left part} that considers an empty graph and the second
term\hide{the adjustment part} that accounts for the edges that actually
occurred in the network:
%
\begin{small}
\begin{align}
  & \nabla_{\Theta_{l}} \LTH =
  \sum_{i, j} \nabla_{\Theta_{l}} \EXATTR{i, j}{\log P(0 |F_{i}, F_{j}, \Theta)} \nonumber \\
  & \quad + \sum_{A_{ij} = 1} \nabla_{\Theta_{l}} \EXATTR{i, j}{\log P(1 |F_{i}, F_{j}, \Theta) - \log P(0 |F_{i}, F_{j}, \Theta)} \,.
  \label{eq:fastmstep1}
\end{align}
\end{small}
Now we approximate the first term that computes the gradient pretending that
the graph $A$ has no edges:
\begin{align}
  & \sum_{i, j} \nabla_{\Theta_{l}} \EXATTR{i, j}{\log P(0 | F_{i}, F_{j}, \Theta)} \nonumber \\
  & = \nabla_{\Theta_{l}} \mathbb{E}_{Q_{i, j}}[\sum_{i, j} \log P(0 | F_{i}, F_{j}, \Theta)] \nonumber \\
  & \approx \nabla_{\Theta_{l}} \EXATTR{i, j}{N(N-1) \mathbb{E}_{F} [\log P(0 | F, \Theta)]} \nonumber \\
  & = \nabla_{\Theta_{l}} N(N-1) \mathbb{E}_{F} [\log P(0 | F, \Theta)] \,.
  \label{eq:fastmstep2}
\end{align}
Since each $F_{il}$ follows the Bernoulli distribution with parameter
$\mu_{l}$, \EQ{eq:fastmstep2} can be
\rev{computed in $O(L)$ time.}
As the second term in \EQ{eq:fastmstep1} requires only $O(LE)$ time, the computation
time of the M-step is reduced from $O(LN^2)$ to
\rev{$O(LE)$.}
Similarly we reduce the computation time of the E-step from $O(L^2 N^2)$ to
$O(L^2 E)$
(see Appendix for details).
Thus overall we reduce the computation time of \MAGFIT from $O(L^2 N^2)$
\rev{to $O(L^2 E)$.}

\section{Experiments}
\label{sec:experiments}
Having introduced the MAG model estimation procedure \MAGFIT, we now turn our
attention to evaluating the fitting procedure itself and the ability of the MAG
model to capture the connectivity structure of real networks. There are three
goals \rev{of} our experiments: (1) evaluate the success of \MAGFIT parameter
estimation procedure; (2) given a network, infer both latent node attributes
and the affinity matrices to accurately model the network structure; (3) given
a network where nodes already have attributes, infer the affinity matrices.
For each experiment, we proceed by describing the experimental setup and
datasets.

\begin{figure}[t]
\centering
\subfigure[Convergence of $\mathcal{L}_{Q}(\mu, \Theta)$\label{fig:convll}]{\includegraphics[width=0.235\textwidth]{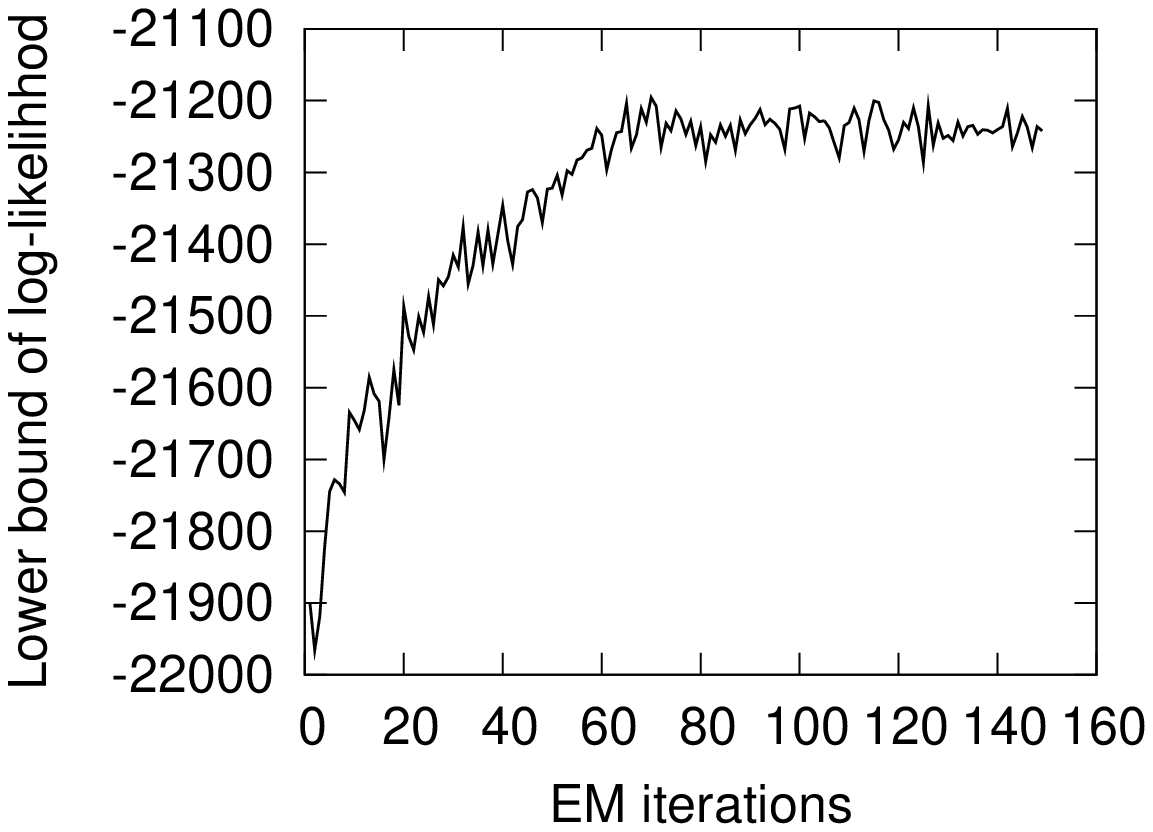}}
\subfigure[Convergence of $\mu_{l}$'s \label{fig:convmu}]{\includegraphics[width=0.235\textwidth]{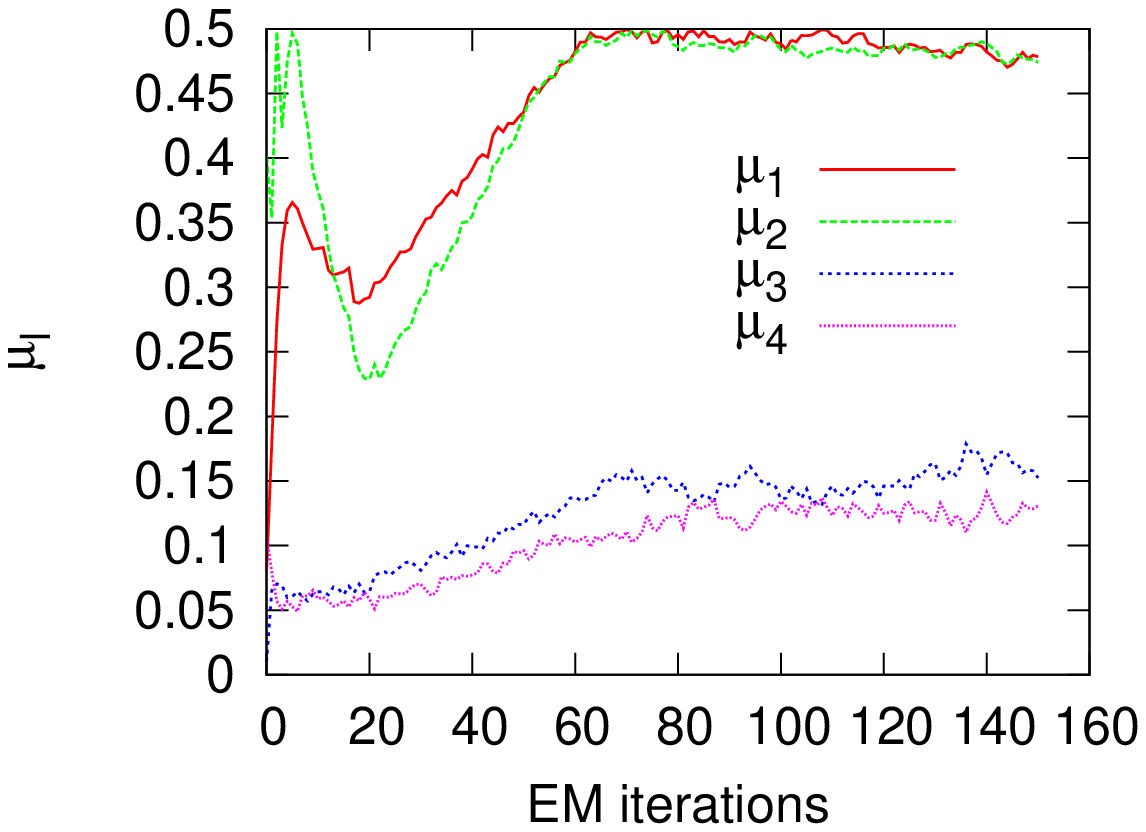}}
\subfigure[Convergence of $\Theta_{l}{[0, 0]}$'s\label{fig:convtheta}]{\includegraphics[width=0.235\textwidth]{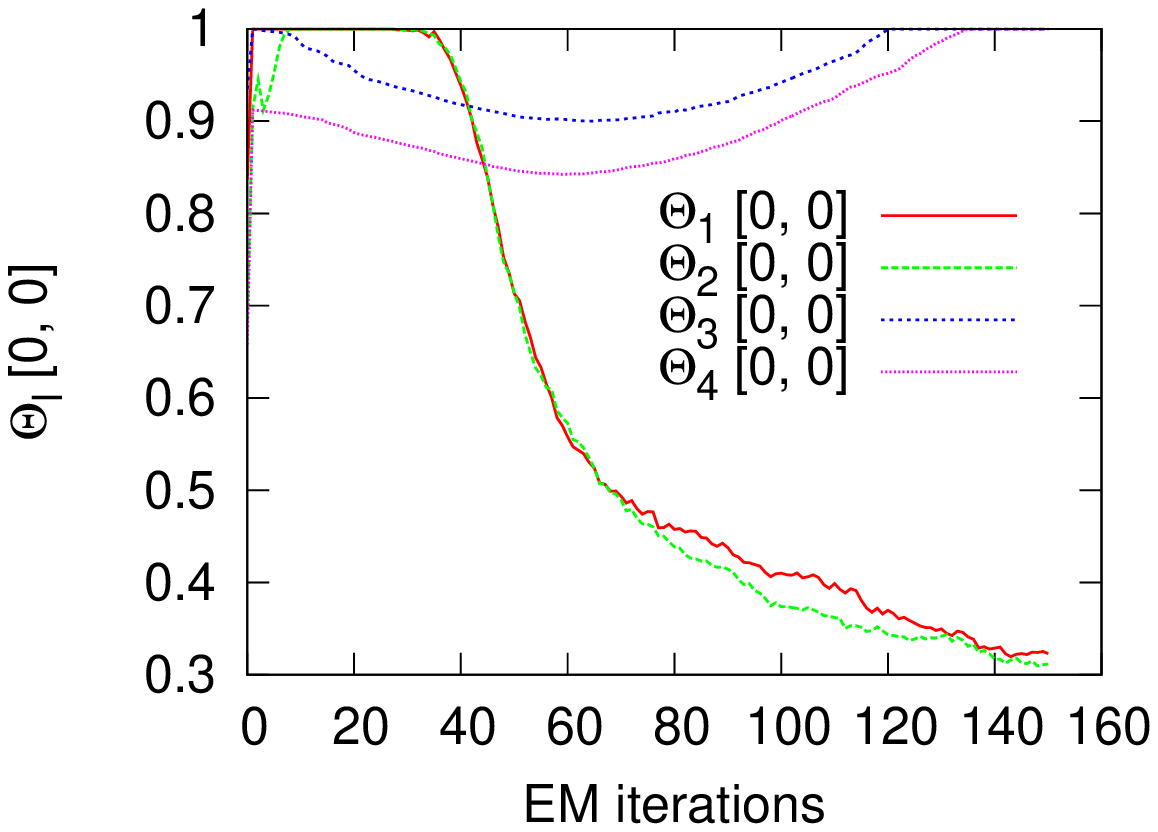}}
\subfigure[\rev{Run time}\label{fig:scale}]{\includegraphics[width=0.235\textwidth]{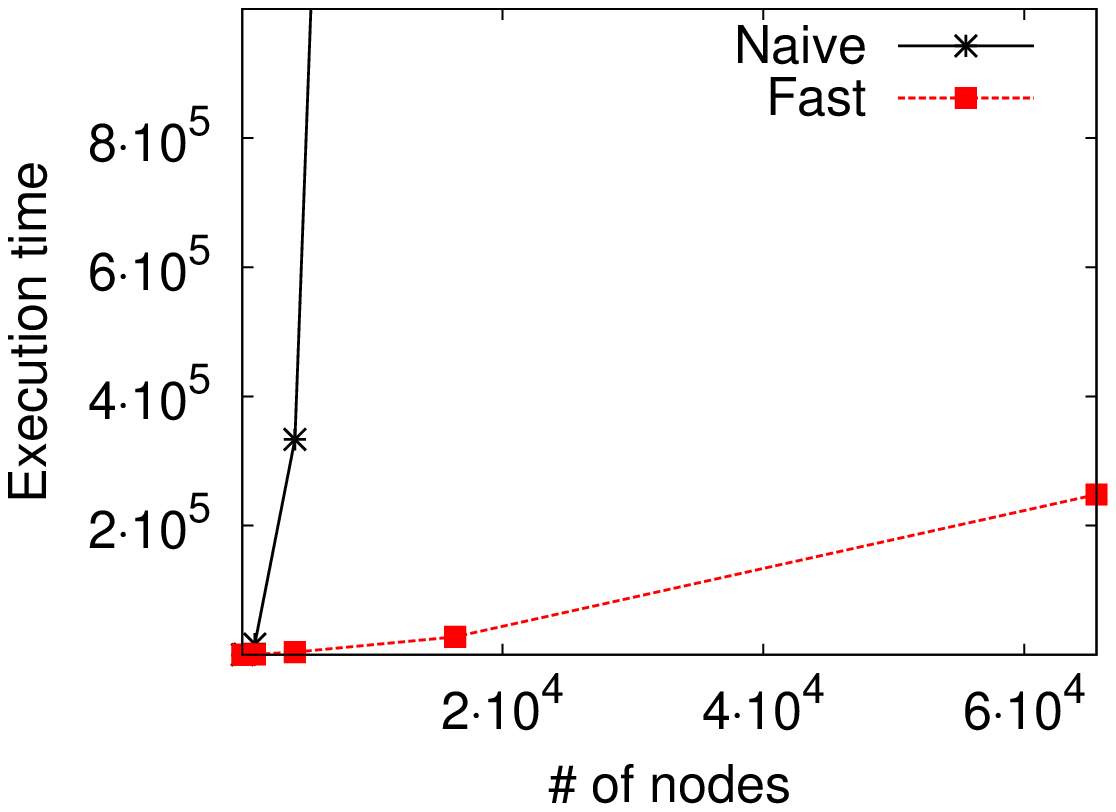}}
\vspace{-3mm}
\caption{{Parameter convergence and scalability.}}
\vspace{-5mm}
\end{figure}

\xhdr{Convergence of \MAGFIT}
First, we briefly evaluate the convergence of the \MAGFIT algorithm. For this
experiment, we use synthetic MAG networks with $N = 1024$ and $L = 4$.
Figure~\ref{fig:convll} illustrates that the objective function $\LQ$, \ie, the
lower bound of the log-likelihood, nicely converges with the number of EM
iterations. While the log-likelihood converges, the model parameters $\mu$ and
$\Theta$ also nicely converge. Figure~\ref{fig:convmu} shows convergence of
$\mu_1, \dots, \mu_4$, while Fig. \ref{fig:convtheta} shows the convergence of
entries $\Theta_l[0,0]$ for $l=1,\dots,4$. Generally, in 100 iterations of EM,
we obtain stable parameter estimates.

We also compare the runtime of the fast \MAGFIT to the naive version where we
do not use speedups for the 
\rev{algorithm.}
Figure~\ref{fig:scale} shows the
runtime as a function of the number of nodes in the network.
%
%
The runtime of the naive algorithm scales quadratically $O(N^2)$, while
the fast version runs in near-linear time.
\rev{For example, on 4,000 node network, the fast algorithm runs about 100 times faster than the naive one.}

Based on these experiments, we conclude that the variational EM gives robust
parameter estimates. We note that the \MAGFIT optimization problem is
non-convex, however, in practice we observe fast convergence and good fits.
Depending on the initialization \MAGFIT may converge to different solutions but
in practice solutions tend to have comparable log-likelihoods and consistently
good fits. Also, the method nicely scales to networks with up to hundred
thousand nodes.

\xhdr{Experiments on real data}
We proceed with experiments on real datasets. We use the LinkedIn social
network~\cite{jure08microevol} at the time in its evolution when it had $N =$
4,096 nodes and $E =$ 10,052 edges.
We also use the Yahoo!-Answers question answering social network, again from
the time when the network had $N =$ 4,096, $E =$ 5,678~\cite{jure08microevol}.
For our experiments we choose $L = 11$, which is roughly $\log N$ as it has
been shown that this is the optimal choice for $L$~\cite{mh10mag}.

Now we proceed as follows. Given a real network $A$, we apply \MAGFIT to
estimate MAG model parameters $\hat{\Theta}$ and $\hat{\mu}$. Then, given these
parameters, we generate a synthetic network $\hat{A}$ and compare how well
synthetic $\hat{A}$ mimics the real network $A$.

\xhdr{Evaluation} To measure the level of agreement between synthetic $\hat{A}$
and the real $A$, we use several different metrics. First, we evaluate how well
$\hat{A}$ captures the structural properties, like degree distribution and
clustering coefficient, of the real network $A$. We consider the following
network properties:

\vspace{-3mm}
\begin{itemize}
  \denselist
  \item {\em In/Out-degree distribution (InD/OutD)} is a histogram of the
      number of in-coming and out-going links of a node.
  \item {\em Singular values (SVal)} indicate the singular values of the
      adjacency matrix versus their rank.
  \item {\em Singular vector (SVec)} represents the distribution of
      components in the left singular vector associated with the largest
      singular value.
  \item {\em Clustering coefficient (CCF)} represents the degree versus the
      average (local) clustering coefficient of nodes of a given
      degree~\cite{watts98smallworld}.
  \item {\em Triad participation (TP)} indicates the number of triangles
      that a node is adjacent to. It measures the transitivity in networks.
\end{itemize}
\vspace{-3mm}

Since distributions of the above quantities are generally heavy-tailed, we plot
them in terms of complementary cumulative distribution functions ($P(X>x)$ as a
function of $x$). Also, to indicate the scale, we do not normalize the
distributions to sum to 1.


\newcommand{\KS}{\textit{KS}}
\newcommand{\POWL}{\textit{L2}}

Second, to quantify the discrepancy of network properties between real and
synthetic networks, we use a variant of Kolmogorov-Sminorv (KS) statistic and
the $L2$ distance between different distributions.
The original KS statistics is not appropriate here since if the distribution
follows a power-law then the original KS statistics is usually dominated by the
head of the distribution. We thus consider the following variant of the KS
statistic: $\KS(D_{1}, D_{2}) = \max_{x} | \log D_{1}(x) - \log
D_{2}(x)|$~\cite{mh11kronem}, where $D_{1}$ and $D_{2}$ are two complementary
cumulative distribution functions.
Similarly, we also define a variant of the $L2$ distance on the log-log scale,
$\POWL(D_{1}, D_{2}) = \sqrt{\frac{1}{\log b - \log a} \left( \int_{a}^{b}
\left(\log D_{1}(x) - \log D_{2}(x)\right)^{2} \, d (\log x)\right)} $ where
$[a, b]$ is the support of distributions ${D_{1}}$ and ${D_{2}}$. Therefore, we
evaluate the performance with regard to the recovery of the network properties
in terms of the $\KS$ and $\POWL$ statistics.

\newcommand{\TPI}{\textit{TPI}\xspace}
\newcommand{\LLX}{\textit{LL}\xspace}

Last, since MAG generates a probabilistic adjacency matrix $P$, we also
evaluate how well $P$ represents a given network $A$. We use the following two
metrics:

\vspace{-3mm}
\begin{itemize}
  \denselist
  \item {\em Log-likelihood (\textit{LL})} measures the possibility that
      the probabilistic adjacency matrix $P$ generates network $A$: $LL=
      \sum_{i j} \log (P_{ij}^{A_{ij}} (1-P_{ij})^{1-A_{ij}})$.
  \item {\em True Positive Rate Improvement (\TPI)} represents the
      improvement of the true positive rate over a random graph: $TPI =
      \sum_{A_{ij}=1}P_{ij} / \frac{E^2}{N^2}$. \TPI\xspace indicates how
      much more probability mass is put on the edges compared to a random
      graph (where each edge occurs with probability $E/N^2$).
\end{itemize}
\vspace{-3mm}


\begin{figure}[t]
  \centering
  \subfigure[In-degree]{\includegraphics[width=0.235\textwidth]{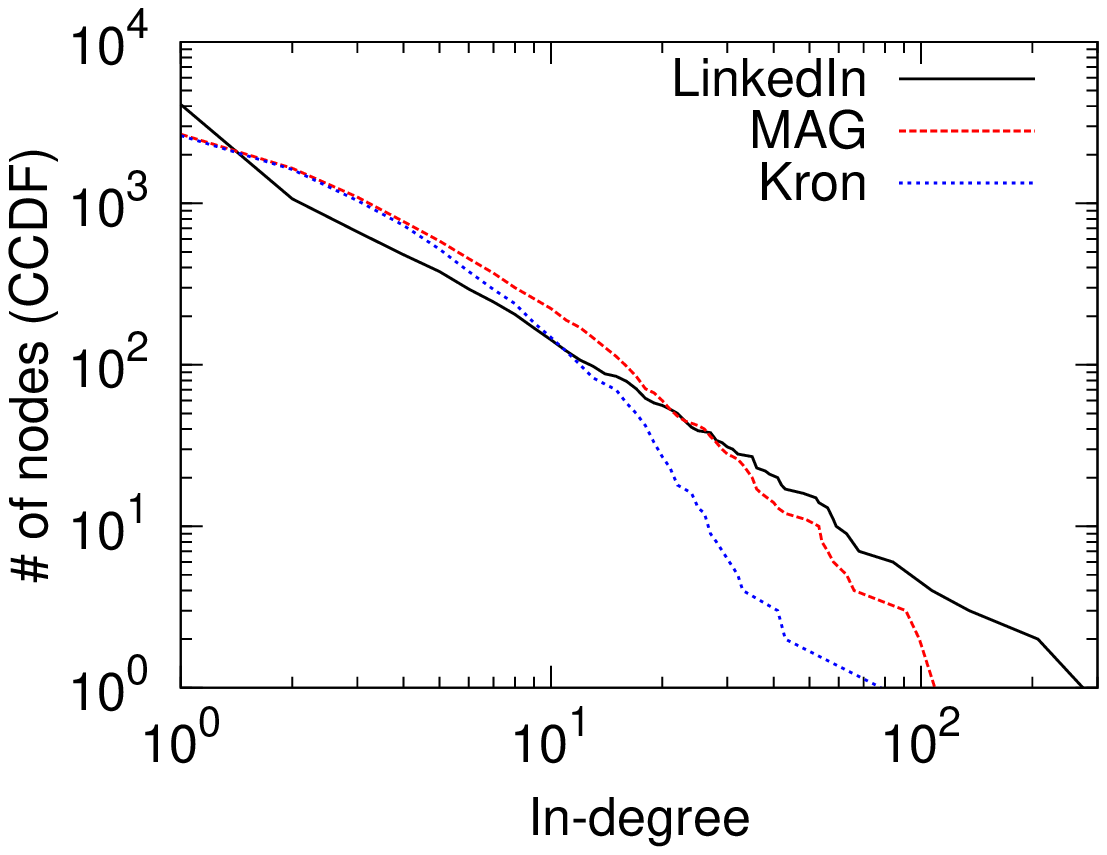}}
  \subfigure[Out-degree]{\includegraphics[width=0.235\textwidth]{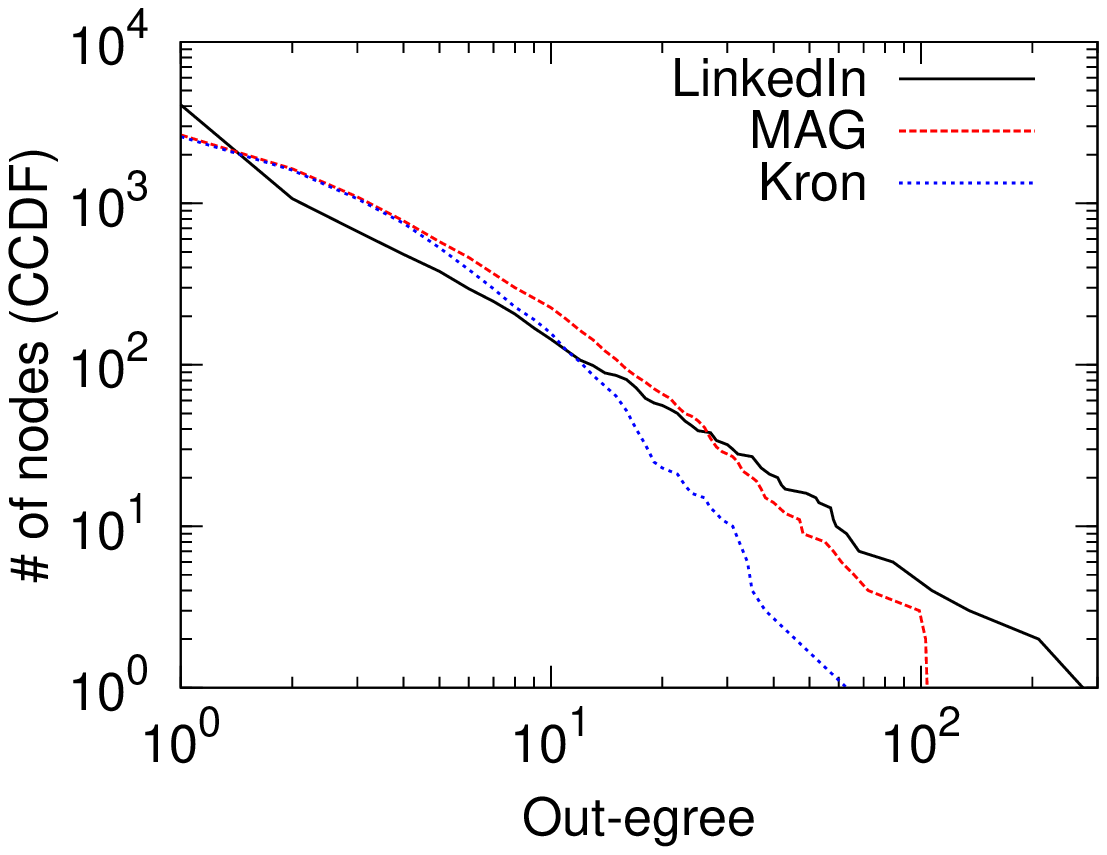}}
  \subfigure[Singular value]{\includegraphics[width=0.235\textwidth]{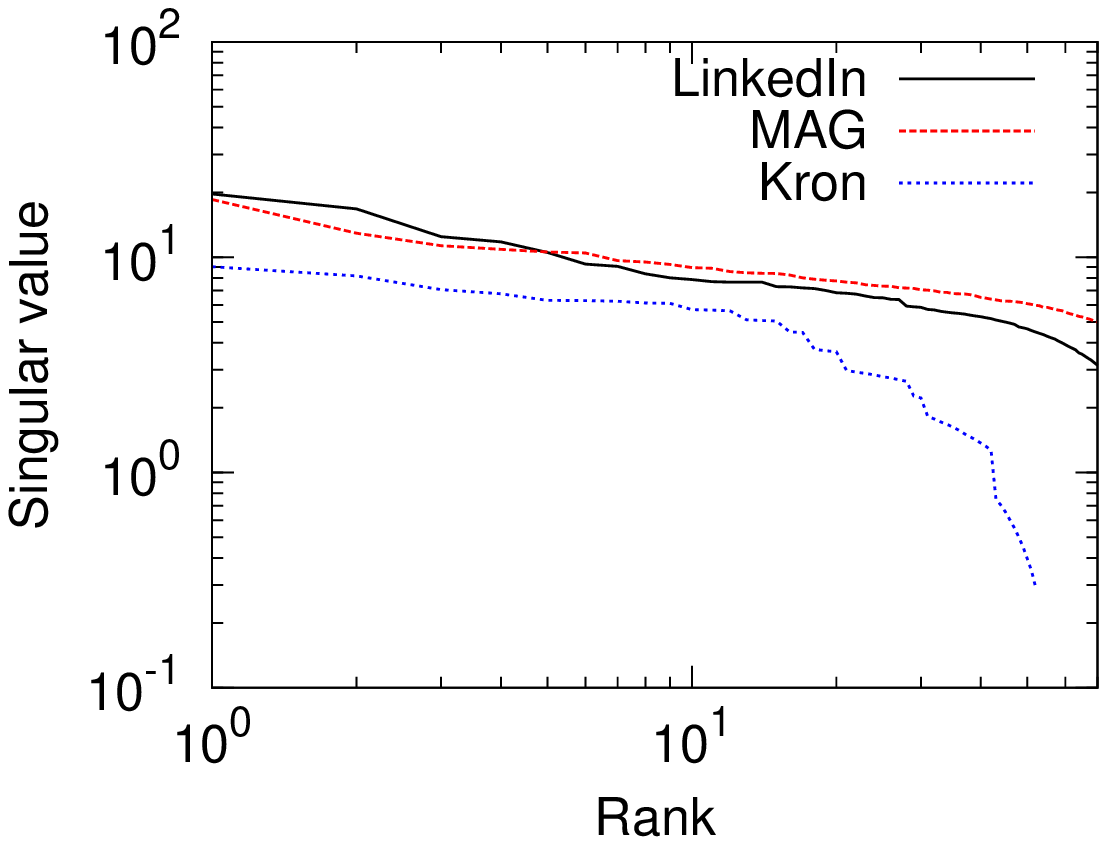}}
  \subfigure[Singular vector]{\includegraphics[width=0.235\textwidth]{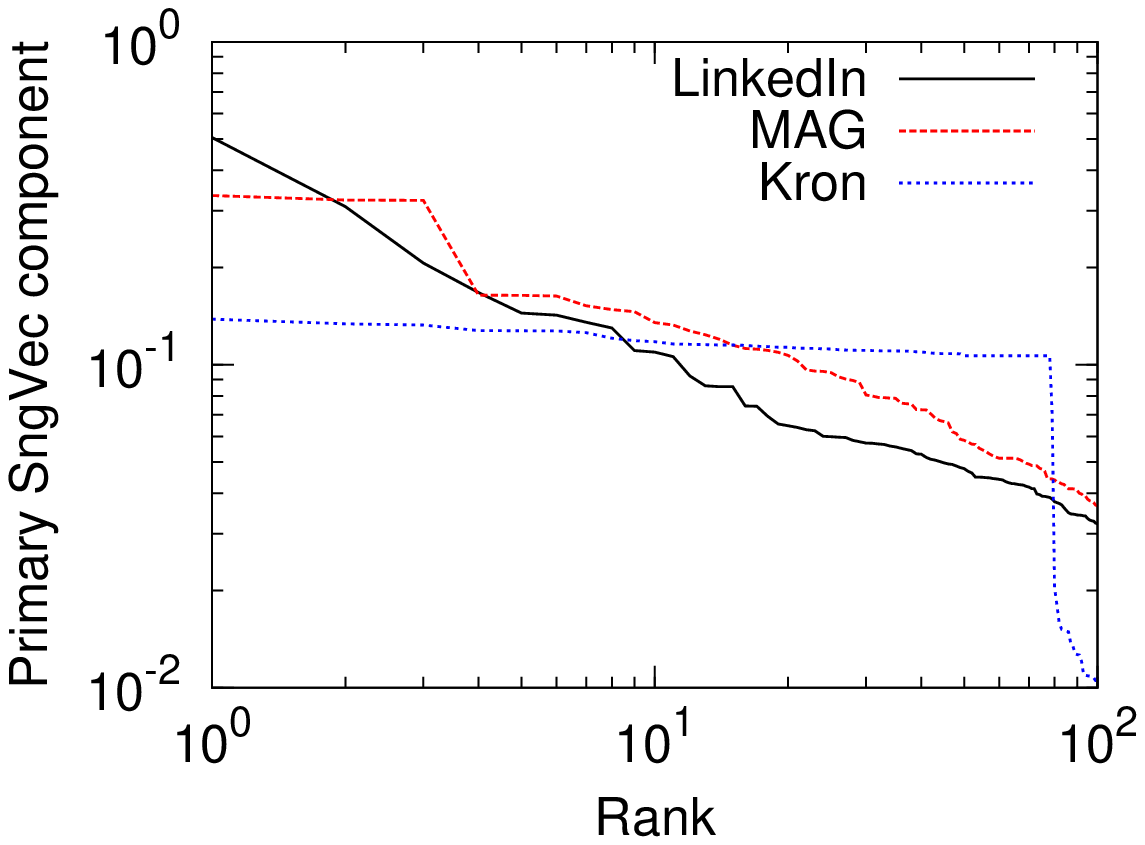}}
  \subfigure[Clustering coefficient]{\includegraphics[width=0.235\textwidth]{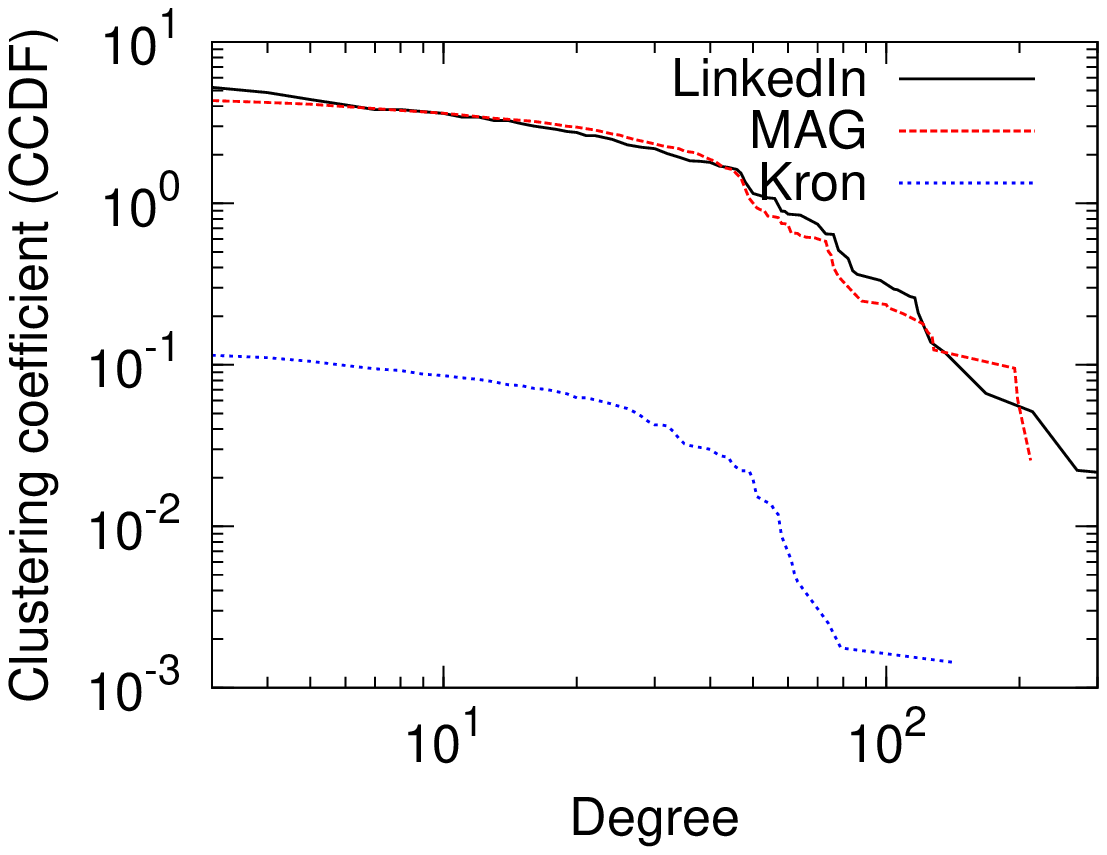}}
  \subfigure[Triad participation]{\includegraphics[width=0.235\textwidth]{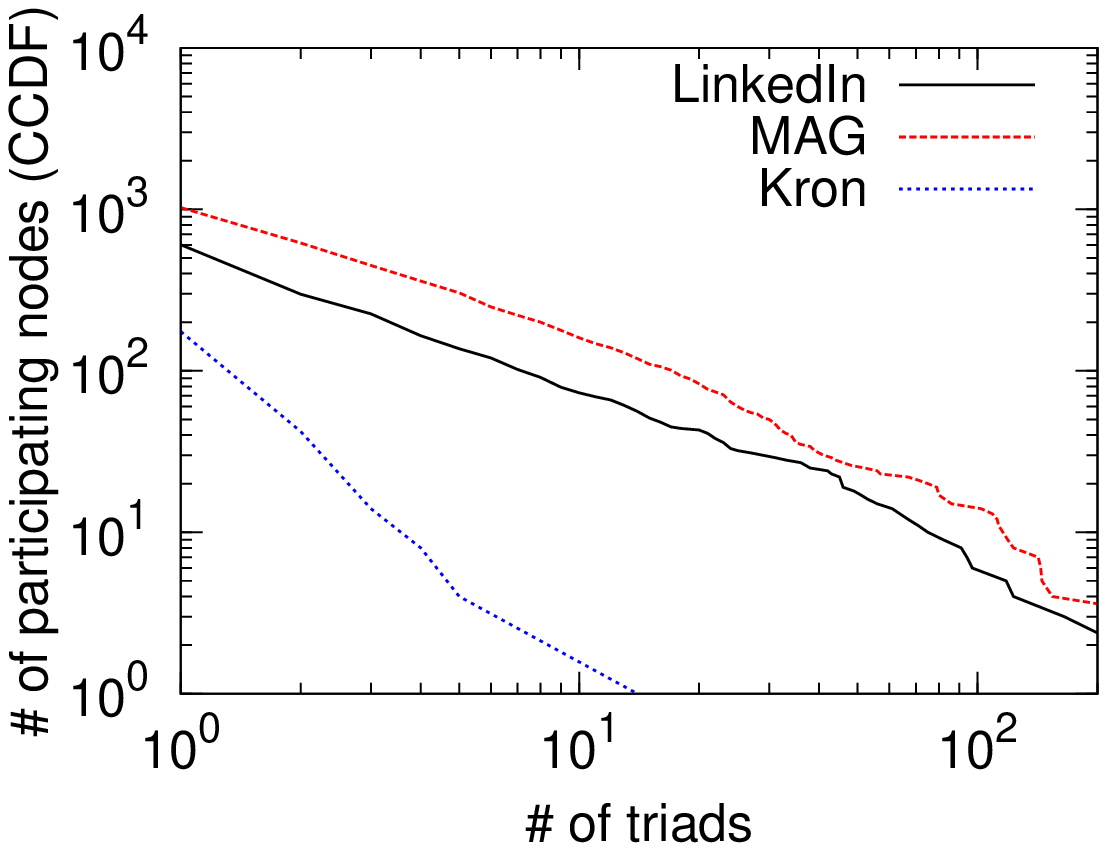}}
  \caption{The recovered network properties by the \MAG~and the Kronecker graphs
  model on the LinkedIn network.
  For every network property, \MAG~outperforms the Kronecker graphs model.}
  \label{fig:realplot}
  \vspace{-3mm}
\end{figure}

\begin{table}[t]
\caption{\KS~and \POWL~of MAG and the Kronecker graphs model on the LinkedIn
network. MAG exhibits 50-70\% better performance than Kronecker graphs model.}
\label{tbl:linkedinks} \centering \small
\begin{tabular}{c||c|c|c|c|c|c||c}
\multicolumn{8}{l}{\small{}}\\
  {\bf \KS} & InD & OutD & SVal & SVec & TP & CCF & Avg \\ \hline \hline
  MAG  & \small 3.70 & 3.80 & 0.84 & 2.43 & 3.87 & 3.16 & 2.97 \\ \hline
  Kron & 4.00 & 4.32 & 1.15 & 7.22 & 8.08 & 6.90 & 5.28 \\
  \multicolumn{8}{l}{\bf \POWL} \\ \hline \hline
  MAG  & 1.01 & 1.15 & 0.46 & 0.62 & 1.68 & 1.11 & 1.00 \\ \hline
  Kron & 1.54 & 1.57 & 0.65 & 6.14 & 6.00 & 4.33 & 3.37
\end{tabular}
\vspace{-3mm}
\end{table}

\xhdr{Recovery of the network structure}
We begin our investigations of real networks by comparing the performance of
the MAG model to that of the Kronecker graphs model~\rev{\cite{jure10kronecker}, which}
offers a state of the art baseline for
modeling the structure of large networks.
We use evaluation methods described in the previous section where \hide{a given
real-world network $A$, }we fit both models to a given real-world network $A$
and generate synthetic $\hat{A}_{MAG}$ and $\hat{A}_{Kron}$. Then we compute
the structural properties of all three networks and plot them in
Figure~\ref{fig:realplot}. Moreover, for each of the properties we also compute
\KS~and \POWL~statistics and show them in Table~\ref{tbl:linkedinks}.

Figure~\ref{fig:realplot} plots the six network properties described above for
the LinkedIn network and the synthetic networks generated by fitting MAG and
Kronecker models to the LinkedIn network.
We observe that MAG can successfully produce synthetic networks that match the
properties of the real network.
In particular, both MAG and Kronecker graphs models capture the degree
distribution of the LinkedIn network well. However, \MAG~performs much better
in matching spectral properties of graph adjacency matrix as well as the local
clustering of the edges in the network.

Table~\ref{tbl:linkedinks} shows the \KS~and \POWL~statistics for each of the
six structural properties plotted in Figure~\ref{fig:realplot}. Results confirm
our previous visual inspection. The \MAG~is able to fit the network structure
much better than the Kronecker graphs model. In terms of the average
\KS~statistics, we observe 43\% improvement, while observe even greater
improvement of 70\% in the \POWL~metric. For degree distributions and the
singular values, MAG outperforms Kronecker for about 25\% while the improvement
on singular vector, triad participation and clustering coefficient is 60 $\sim$
75\%.

We make similar observations on the Yahoo!-Answers network but omit the results
for brevity. We include them
in Appendix.

\begin{figure}[t]
\centering
\begin{tabular}{ccc}
  \includegraphics[width=0.15\textwidth]{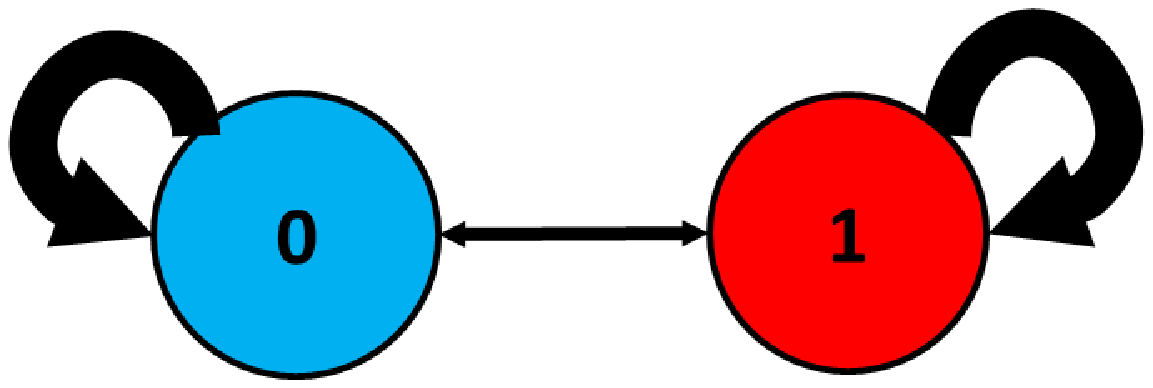} &
  \includegraphics[width=0.13\textwidth]{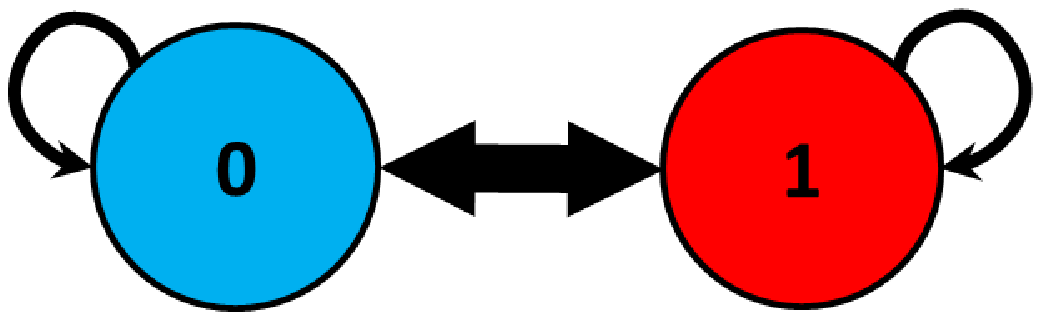} &
  \includegraphics[width=0.13\textwidth]{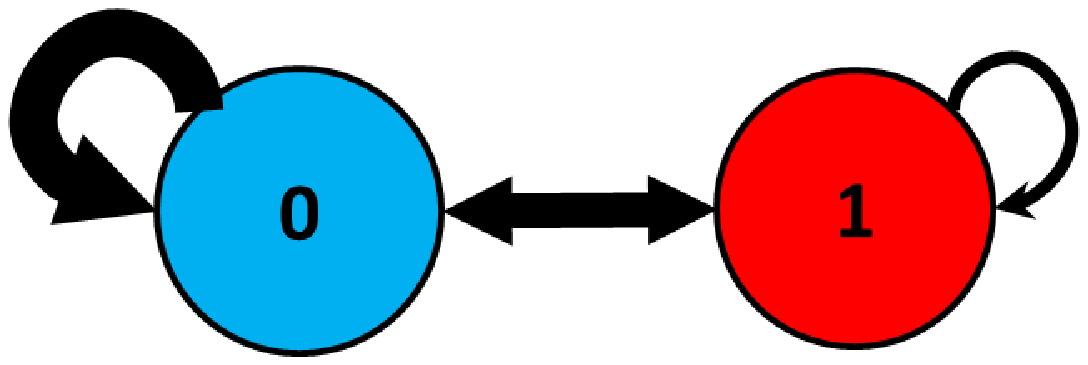} \\
  \includegraphics[width=0.1\textwidth]{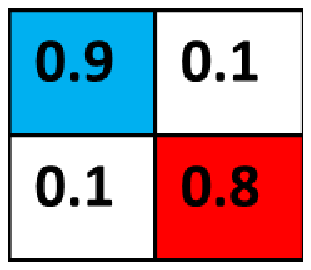} &
  \includegraphics[width=0.1\textwidth]{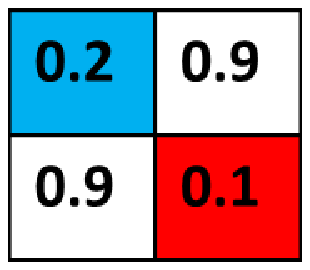} &
  \includegraphics[width=0.1\textwidth]{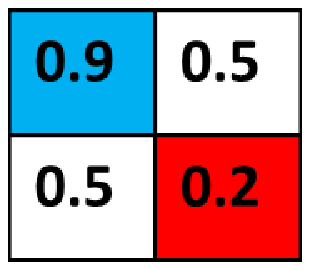} \\
  \small (a) Homophily & \small (b) Heterophily & \small (c) Core-Periphery \\
\end{tabular}
  \caption{Structures in which a node attribute can affect link affinity.
  The widths of arrows
  correspond to the affinities towards link formation.}
  \label{fig:structure}
\vspace{-3mm}
\end{figure}

We interpret the improvement of the MAG over Kronecker graphs model in the
following way. Intuitively, we can think of Kronecker graphs model as a version
of the MAG model where all affinity matrices $\Theta_l$ are the same and all
$\mu_l=0.5$.
However, real-world networks may include various types of structures and thus
different attributes may interact in different ways. For example,
Figure~\ref{fig:structure} shows three possible linking affinities of a binary
attribute. Figure~\ref{fig:structure}(a) shows a homophily (love of the same)
attribute affinity and the corresponding affinity matrix $\Theta$. Notice large
values on the diagonal entries of $\Theta$, which means that link probability
is high when nodes share the same attribute value.
\rev{The top of each figure}
demonstrates that there will be many links between nodes that have the value of
the attribute set to ``0'' and many links between nodes that have the value
``1'', but there will be few links between nodes where one has value ``0'' and
the other ``1''.
Similarly, Figure~\ref{fig:structure}(b) shows a heterophily (love of the
different) affinity, where nodes that do not share the value of the attribute
are more likely to link, which gives rise to near-bipartite networks. Last,
Figure~\ref{fig:structure}(c) shows a core-periphery affinity, where links are
most likely to form between ``0'' nodes (\ie, members of the core) and least
likely to form between ``1'' nodes (\ie, members of the periphery). Notice that
links between the core and the periphery are more likely than the links between
the nodes of the periphery.

%
%
Turning our attention back to MAG and Kronecker models, we note that
real-world networks globally exhibit nested core-periphery
structure~\cite{jure10kronecker} (Figure \ref{fig:structure}(c)). While there
exists the core (densely connected) and the periphery (sparsely connected) part
of the network, there is another level of core-periphery structure inside the
core itself. On the other hand, if viewing the network more finely, we may also
observe the homophily which produces local community structure. MAG can model
both global core-periphery structure and local homophily communities, while the
Kronecker graphs model cannot express the different affinity types because it
uses only one initiator matrix.



For example, the LinkedIn network consists of 4 core-periphery affinities, 6
homophily affinities, and 1 heterophily affinity matrix. Core-periphery
affinity models active users who are more likely to connect to others.
Homophily affinities model people who are more likely to connect to others in
the same job area. Interestingly, there is a heterophily affinity which results
in bipartite relationship. We believe that the relationships between job
seekers and recruiters or between employers and employees leads to this
structure.

\begin{table}[t]
  \caption{{\textit{LL} and \TPI values for LinkedIn (\textit{LI}) and
  Yahoo!-Answers (\textit{YA}) networks}} \label{tbl:linkacc}
\centering \small
  \vspace{-2mm}
\begin{tabular}{c||c|c||c|c}
\multicolumn{5}{l}{}\\
  & \textit{LL}(\textit{LI}) & \TPI(\textit{LI}) & \textit{LL}(\textit{YA}) & \TPI(\textit{YA}) \\ \hline \hline
  MAG & -47663 & 232.8 & -33795 & 192.2 \\ \hline
  Kron & -87520 & 10.0 & -48204 & 5.4
\end{tabular}
  \vspace{-3mm}
\end{table}

\xhdr{TPI and LL}
We also compare the \textit{LL} and \TPI~values of MAG and Kronecker models on
both LinkedIn and Yahoo!-Answers networks. Table~\ref{tbl:linkacc} shows that
MAG outperforms Kronecker graphs by surprisingly large margin.
\rev{In \textit{LL} metric, the MAG model shows $50\sim\,60$ \% improvement over the Kronecker model.}
Furthermore, in \TPI~metric, the MAG model shows $23\sim\,35$ times better
accuracy than the Kronecker model. From these results, we conclude that the
\MAG~achieves a superior probabilistic representation of a given network.


%
%
\xhdr{Case Study: \textit{AddHealth} network}
So far we considered node attributes as {\em latent} and we inferred the
affinity matrices $\Theta$ as well as the attributes themselves. Now, we
consider the setting where the node attributes are already given and we only
need to infer affinities $\Theta$. Our goal here is to study how real
attributes explain the underlying network structure.

We use the largest high-school friendship network ($N =$ 457, $E =$ 2,259) from
the National Longitudinal Study of Adolescent Health (\textit{AddHealth})
dataset. The dataset includes more than 70 school-related attributes for each
student. Since some attributes do not take binary values, we binarize them
by taking value 1 if the value of the attribute is less than the
median value. Now we aim to investigate which attributes affect the friendship
formation and how.


We set $L=7$ and consider the following methods for selecting a subset of 7
attributes:

\vspace{-3mm}
\begin{itemize}
  \denselist
  \item {\em R7}: Randomly choose 7 real attributes and fit the model (\ie,
      only fit $\Theta$ as attributes are given).
  \item {\em L7}: Regard all 7 attributes as latent (\ie, not given) and
      estimate $\mu_l$ and $\Theta_l$ for $l=1,\dots,7$.
  \item {\em F7}: Forward selection. Select attributes one by one. At each
      step select an additional attribute that maximizes the overall
      log-likelihood (\ie, select a real attribute and estimate its
      $\Theta_l$).
  \item {\em F5+L2}: Select 5 real attributes using forward selection.
      Then, we infer 2 more latent attributes.
\end{itemize}
\vspace{-3mm}


To make the \MAGFIT work with fixed real attributes (\ie, only infer $\Theta$)
we fix $\phi_{il}$ to the values of real attributes. In the E-step we
then optimize only over the latent set of $\phi_{il}$ and the M-step remains as
is.

\begin{figure}[t]
\centering
\includegraphics[width=0.235\textwidth]{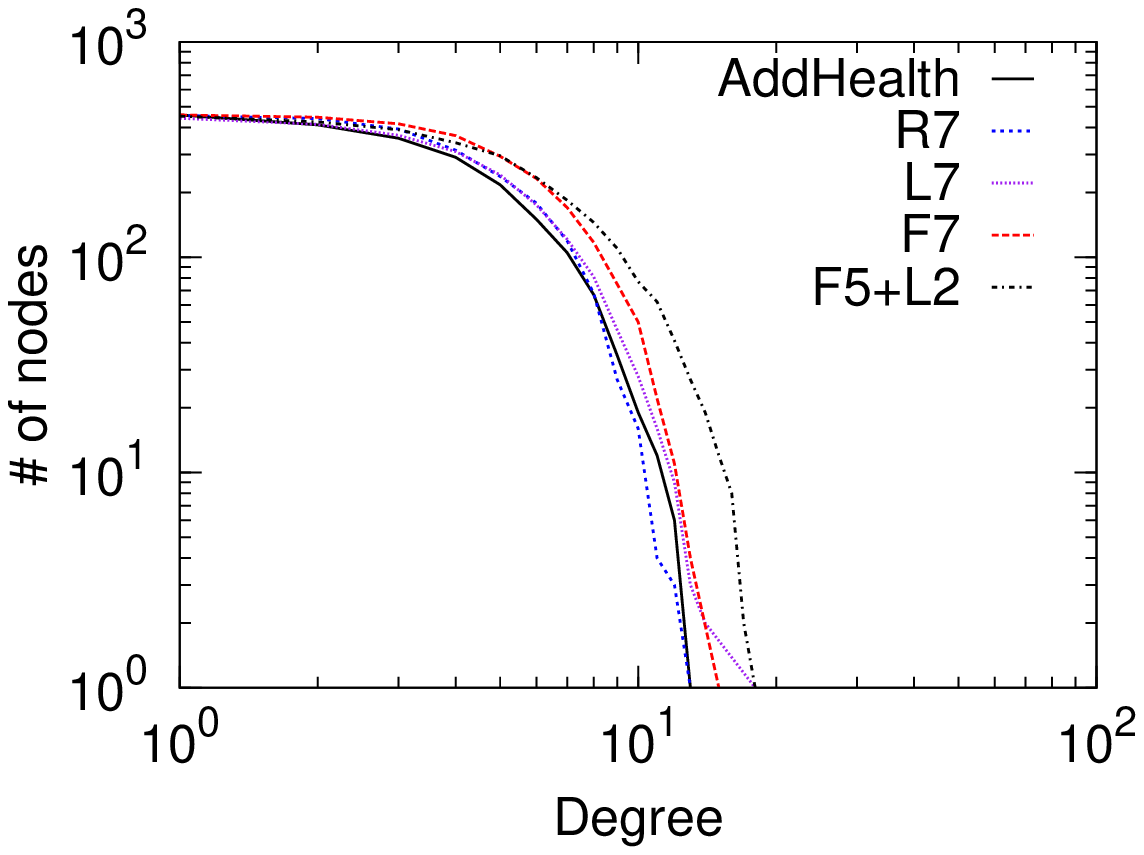}
\includegraphics[width=0.235\textwidth]{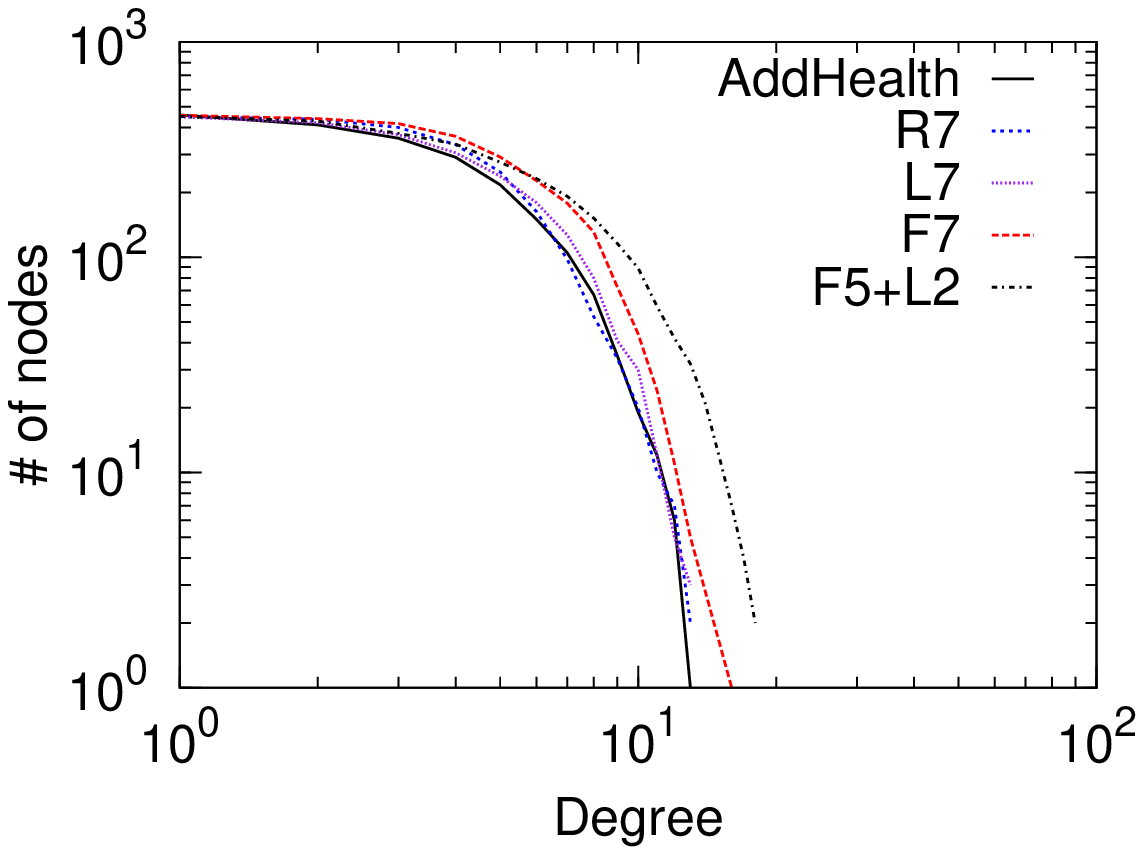}
\includegraphics[width=0.235\textwidth]{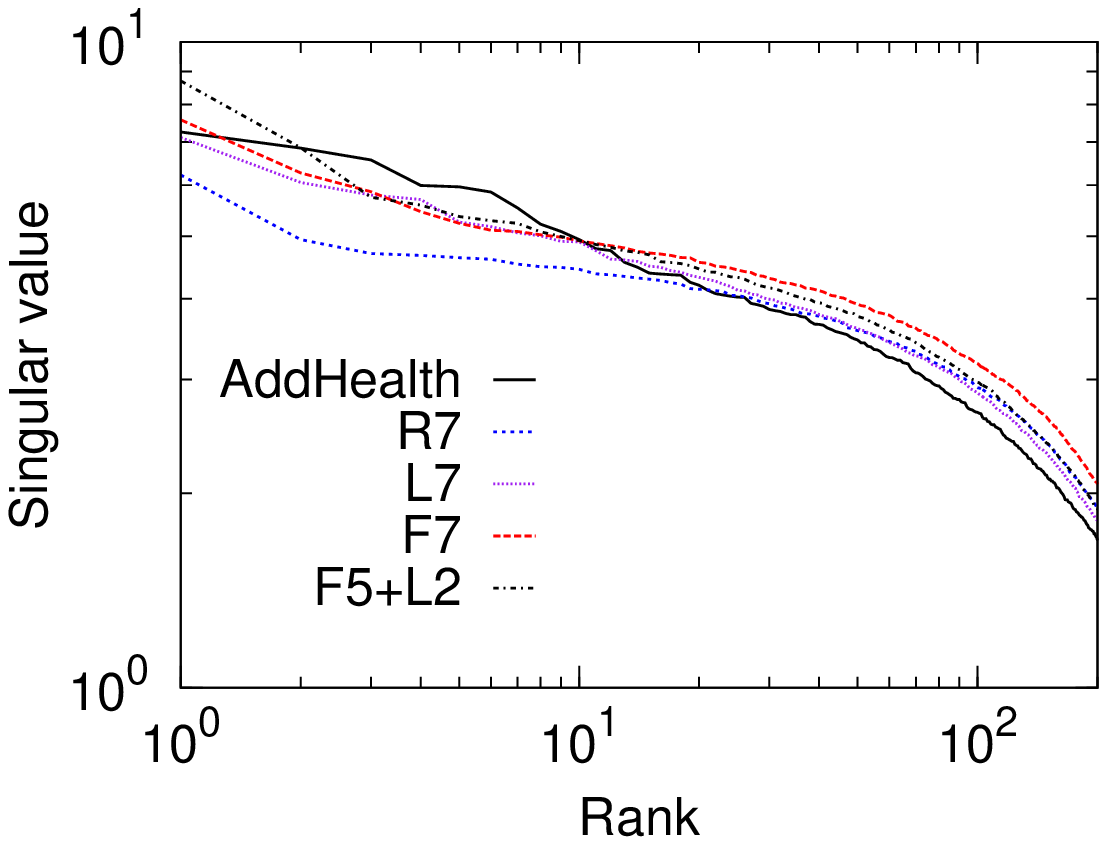}
\includegraphics[width=0.235\textwidth]{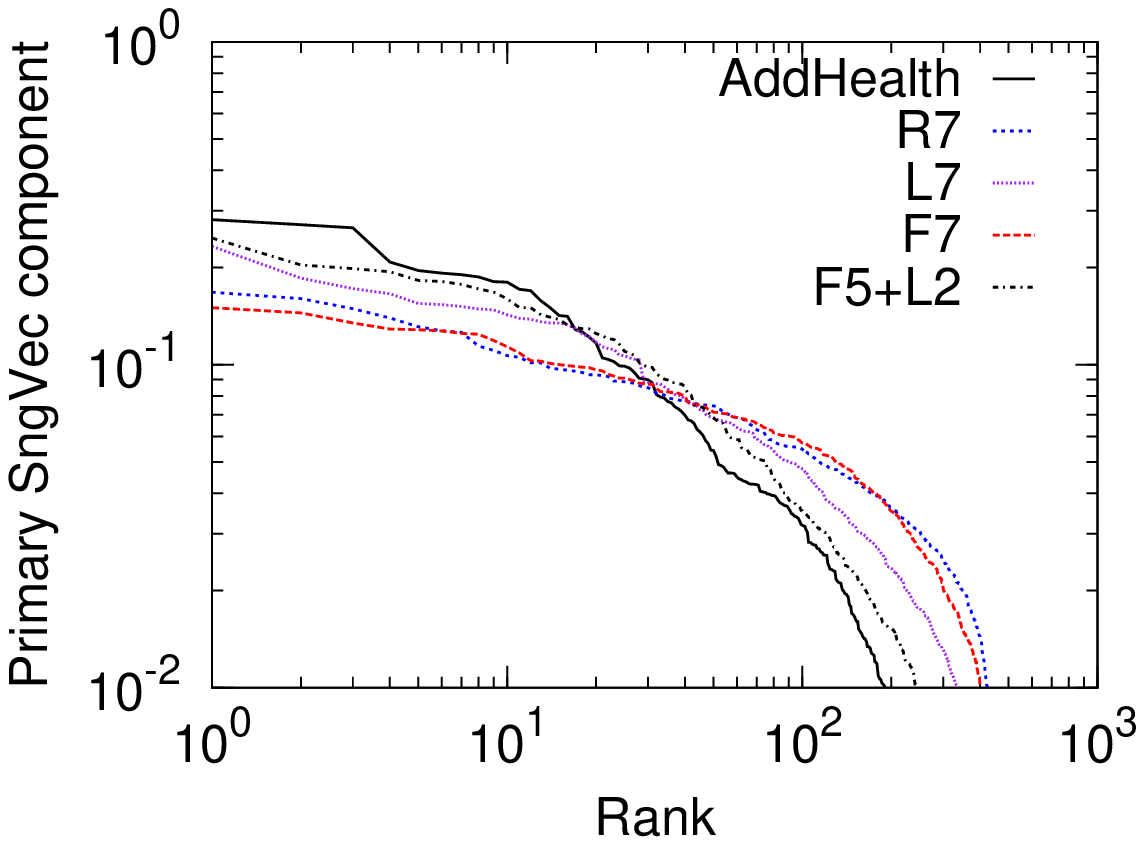}
\includegraphics[width=0.235\textwidth]{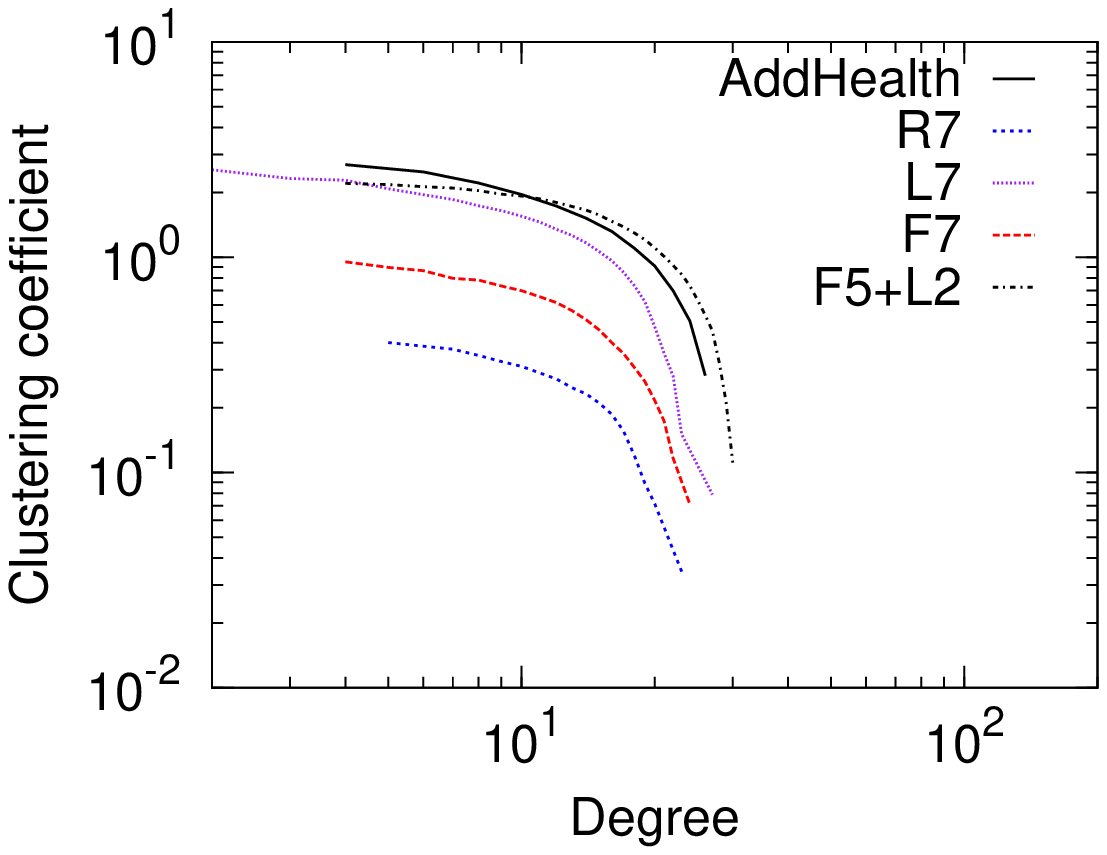}
\includegraphics[width=0.235\textwidth]{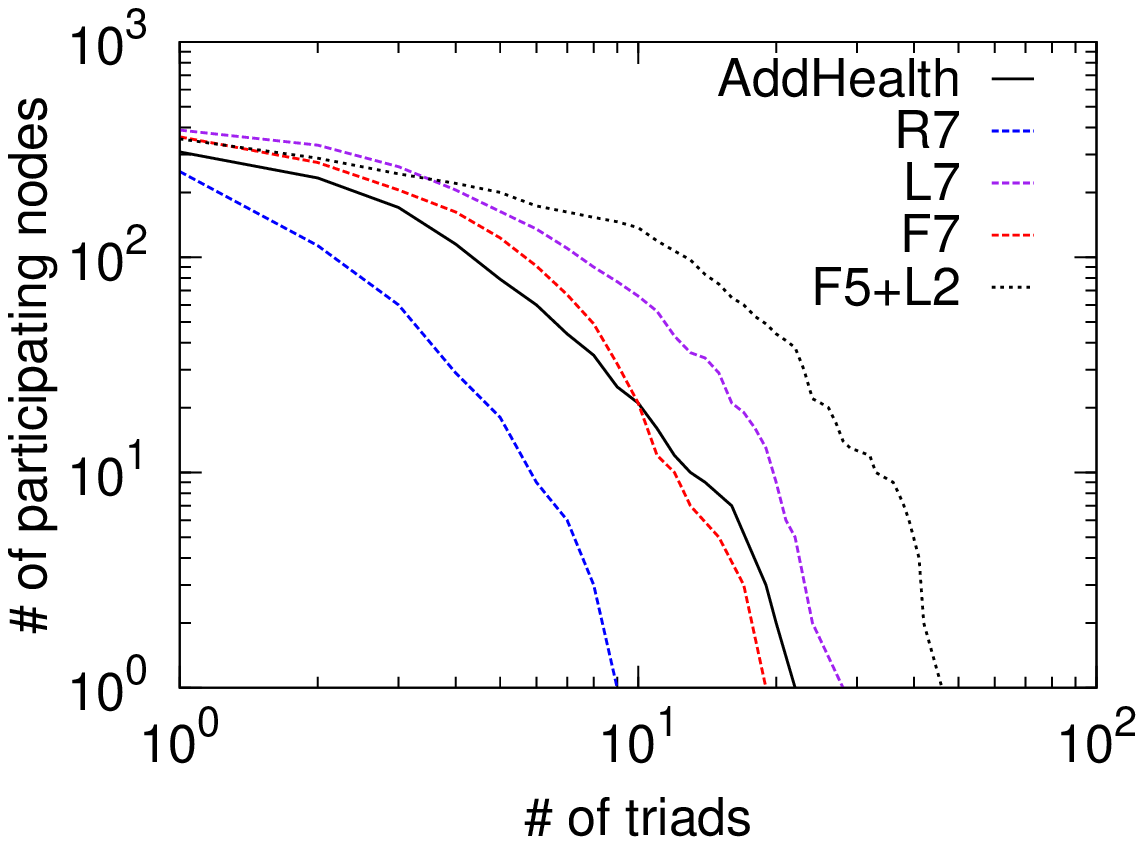}
\vspace{-8mm}
\caption{{Properties of the AddHealth network.}}
\vspace{-5mm}
\label{fig:addhealth}
\end{figure}

\xhdr{{\em AddHealth} network structure}
We begin by evaluating the recovery of the network structure.
Figure~\ref{fig:addhealth} shows the recovery of six network properties for
each attribute selection method. We note that each method manages to recover
degree distributions as well as spectral properties (singular values and
singular vectors) but the performance is different for clustering coefficient
and triad participation.

\begin{table}[t]
\centering \small
  \caption{Performance of different selection methods.}
  \label{tbl:addhealthks}
\begin{tabular}{l|c|c|c|c|c|c||c}
\multicolumn{8}{l}{\small{}}\\
  {\bf \KS} & InD & OutD & SVal & SVec & TP & CCF & Avg \\ \hline \hline
  R7  &1.00 & 0.58 & 0.48 & 2.92 & 4.52 & 4.45 & 2.32 \\ \hline
  F7  &2.32 & 2.80 & 0.30 & 2.68 & 2.60 & 1.58 & 2.05 \\ \hline
  F5+L2&3.45 & 4.00 & 0.26 & 0.95 & 1.30 & 3.45 & 2.24 \\ \hline
  L7 &1.58 & 1.58 & 0.18 & 2.00 & 2.67 & 2.66 & 1.78 \\
  \multicolumn{8}{l}{\bf \POWL} \\ \hline \hline
  R7   &0.25 & 0.16 & 0.25 & 0.96 & 3.18 & 1.74 & 1.09 \\ \hline
  F7  &0.71 & 0.67 & 0.18 & 0.98 & 1.26 & 0.78 & 0.76 \\ \hline
  F5+L2&0.80 & 0.87 & 0.13 & 0.34 & 0.76 & 1.30 & 0.70 \\ \hline
  L7 &0.29 & 0.27 & 0.10 & 0.64 & 0.75 & 1.22 & 0.54
\end{tabular}
  \vspace{-5mm}
\end{table}

Table~\ref{tbl:addhealthks} shows the discrepancies in the 6 network properties
(\KS~and \POWL~statistics) for each attribute selection method. As expected,
selecting 7 real attributes at random (R7) performs the worst. Naturally, L7
performs the best (23\% improvement over R7 in \KS~and 50\% in \POWL) as it has
the most degrees of freedom\hide{ ($(1+4)\cdot7=35$ parameters)}. It is
followed by F5+L2 (the combination of 5 real and 2 latent attributes) and F7
(forward selection).

As a point of comparison we also experimented with a simple logistic regression
classifier where given the attributes of a pair of nodes we aim to predict an
occurrence of an edge. Basically, given network $A$ on $N$ nodes, we have $N^2$
(one for each pair of nodes) training examples: $E$ are positive (edges) and $N^2-E$ are negative (non-edges).
However, the model performs poorly as it gives 50\% worse \KS~statistics than
MAG. The average \KS~of logistic regression under R7 is 3.24 (vs. 2.32 of MAG) and the same statistic under F7 is 3.00 (vs. 2.05 of MAG). Similarly, logistic regression gives 40\% worse \POWL~under R7
and 50\% worse \POWL~under F7.
These results demonstrate that using the same
attributes MAG heavily outperforms logistic regression.
\rev{We understand that this performance difference arises because the connectivity between a pair of nodes depends on some factors other than the linear combination of their attribute values.}

\begin{table}[t]
  \caption{{\textit{LL} and \TPI~for the {\em AddHealth} network.}}
\label{tbl:addhealthacc}
\centering \small
\begin{tabular}{c|c|c|c|c}
\multicolumn{5}{l}{\small{}}\\
  & R7 & F7 & F5+L2 & \hide{Kron & }L7 \\ \hline \hline
  \textit{LL} & -13651 & -12161 & -12047 & \hide{-11629 &} -9154 \\ \hline
  \TPI &  1.0 & 1.1 & 1.9 & \hide{3.0 &}  10.0
\end{tabular}
\vspace{-5mm}
\end{table}

Last, we also examine the \textit{LL} and \TPI values and compare them to the
random attribute selection R7 as a baseline. Table~\ref{tbl:addhealthacc} gives
the results. Somewhat contrary to our previous observations, we note that F7
only slightly outperforms R7, while F5+L2 gives a factor 2 better \TPI than R7.
Again, L7 gives a factor 10 improvement in \TPI and overall best performance.

\begin{table}[t]
  \caption{{Affinity matrices of 5 AddHealth attributes.}}
  \label{tbl:attrs}
\centering \small \vspace{-2mm}
\begin{tabular}{ l | l }
\multicolumn{2}{l}{\small{}}\\
  \hide{$\mu$ &} Affinity matrix & Attribute description \\ \hline \hline
  \hide{0.624 &} [0.572\,0.146;\,0.146\,0.999] & School year (0 if $\geq$ 2) \\ \hline
  \hide{0.513 &} [0.845\,0.332;\,0.332\,0.816] & Highest level math (0 if $\geq$ 6) \\ \hline
  \hide{0.502 &} [0.788\,0.377;\,0.377\,0.784] & Cumulative GPA (0 if $\geq$ 2.65) \\ \hline
  \hide{0.078 &} [0.999\,0.246;\,0.246\,0.352] & AP/IB English (0 if taken) \\ \hline
  \hide{0.672 &} [0.794\,0.407;\,0.407\,0.717] & Foreign language (0 if taken)
\end{tabular}
\vspace{-5mm}
\end{table}

\xhdr{Attribute affinities}
Last, we investigate the structure of attribute affinity matrices to illustrate
how MAG model can be used to understand the way real attributes interact in
shaping the network structure. We use forward selection (F7) to select 7 real
attributes and estimate their affinity matrices. Table~\ref{tbl:attrs}  reports
first 5 attributes selected by the forward selection.

First notice that {\em AddHealth} network is undirected graph and that the
estimated affinity matrices are all symmetric. This means that without a priori
biasing the fitting towards undirected graphs, the recovered parameters obey
this structure.
Second, we also observe that every attribute forms a homophily structure in a
sense that each student is more likely to be friends with other students of the
same characteristic. For example, people are more likely to make friends of the
same school year. Interestingly, students who are freshmen or sophomore are
more likely (0.99) to form links among themselves than juniors and seniors (0.57).
Also notice that the level of advanced courses that each student takes as well
as the GPA affect the formation of friendship ties. Since it is difficult for
students to interact if they do not take the same courses, the chance of the
friendships may be low. We note that, for example, students that
take advanced placement (AP) English courses are very likely to form links.
However, links between students who did not take AP English are nearly as
likely as links between AP and non-AP students. Last, we also observe
relatively small effect of the number of foreign language courses taken on the
friendship formation.

\section{Conclusion}
\label{sec:conclusion}
We developed \MAGFIT, a scalable variational expectation maximization method
for parameter estimation of the Multiplicative Attribute Graph model. The model
naturally captures interactions between node attributes and the network
structure. MAG model considers nodes with categorical attributes and the
probability of an edge between a pair of nodes depends on the product of
individual attribute link formation affinities.
Experiments show that MAG reliably captures the network connectivity patterns
as well as provides insights into how different attributes shape the structure
of networks.
Venues for future work include settings where node attributes are partially
missing and investigations of other ways to combine individual attribute
linking affinities into a link probability.

\subsubsection*{Acknowledgements}
Research was in-part supported by NSF
CNS-1010921,  
NSF IIS-1016909,    
AFRL FA8650-10-C-7058, 
LLNL DE-AC52-07NA27344, 
Albert Yu \& Mary Bechmann Foundation, IBM,
Lightspeed, Yahoo and the Microsoft Faculty Fellowship.


\begin{thebibliography}{10}
\bibitem{airoldi07blockmodel} E.~M. Airoldi, D.~M. Blei, S.~E. Fienberg, and E.~P. Xing.
\newblock Mixed membership stochastic blockmodels.
\newblock {\em JMLR}, 9:1981--2014, 2007.

\bibitem{bonato10waw} A.~Bonato, J.~Janssen, and P.~Pralat.
\newblock The geometric protean model for on-line social networks.
\newblock In {\em WAW '10}, 2010.

\bibitem{faloutsos99powerlaw} M.~Faloutsos, P.~Faloutsos, and C.~Faloutsos.
\newblock On power-law relationships of the internet topology.
\newblock In {\em SIGCOMM '99}, pages 251--262, 1999.

\bibitem{hoff02latent} P.~Hoff, A.~Raftery, and M.~Handcock.
\newblock Latent space approaches to social network analysis.
\newblock {\em Journal of the American Statistical Association}, 97:1090--1098, 2002.

\bibitem{mh10mag} M.~Kim and J.~Leskovec.
\newblock Multiplicative attribute graph model of real-world networks.
\newblock In {\em WAW '10}, 2010.

\bibitem{mh11kronem} M.~Kim and J.~Leskovec.
\newblock Network completion problem: Inferring missing nodes and edges in networks.
\newblock In {\em SDM '11}, 2011.


\bibitem{kumar00stochastic} R.~Kumar, P.~Raghavan, S.~Rajagopalan, D.~Sivakumar, A.~Tomkins, and E.~Upfal.
\newblock Stochastic models for the web graph.
\newblock In {\em FOCS '00}, page~57, 2000.

\bibitem{jure08microevol} J.~Leskovec, L.~Backstrom, R.~Kumar, and A.~Tomkins.
\newblock Microscopic evolution of social networks.
\newblock In {\em KDD '08}, pages 462--470, 2008.

\bibitem{jure10kronecker} J.~Leskovec, D.~Chakrabarti, J.~Kleinberg, C.~Faloutsos, and Z.~Ghahramani.
\newblock {Kronecker Graphs: An Approach to Modeling Networks}.
\newblock {\em JMLR}, 2010.

\bibitem{jure05kronecker} J.~Leskovec, D.~Chakrabarti, J.~M. Kleinberg, and C.~Faloutsos.
\newblock Realistic, mathematically tractable graph generation and evolution, using kronecker multiplication.
\newblock In {\em PKDD '05}, pages 133--145, 2005.

\bibitem{jure07kronfit} J.~Leskovec and C.~Faloutsos.
\newblock Scalable modeling of real graphs using kronecker multiplication.
\newblock In {\em ICML '07}, 2007.

\bibitem{jure05dpl} J.~Leskovec, J.~M. Kleinberg, and C.~Faloutsos.
\newblock Graphs over time: densification laws, shrinking diameters and possible explanations.
\newblock In {\em KDD '05}, 2005.

\bibitem{mahdian07kronecker} M.~Mahdian and Y.~Xu.
\newblock Stochastic kronecker graphs.
\newblock In {\em WAW '07}, pages 179--186, 2007.

\bibitem{page98pagerank} L.~Page, S.~Brin, R.~Motwani, and T.~Winograd.
\newblock The pagerank citation ranking: Bringing order to the web.
\newblock Technical report, Stanford Dig. Lib. Tech. Proj., 1998.

\bibitem{Palla2010} G.~Palla, L.~Lovasz, and T.~Vicsek.
\newblock Multifractal network generator.
\newblock {\em PNAS}, 107(17):7640--7645, 2010.

\bibitem{robins07egrm} G.~Robins and P.~Pattison and Y.~Kalish and D.~Lusher
\newblock An introduction to exponential random graph (p*) models for social networks.
\newblock {\em Social Networks}, 29(2):173--191, 2007.

\bibitem{sarkar05dynamic} P.~Sarkar and A.~W. Moore.
\newblock Dynamic social network analysis using latent space models.
\newblock {\em SIGKDD Explorations}, 7:31--40, December 2005.

\bibitem{watts98smallworld} D.~J. Watts and S.~H. Strogatz.
\newblock Collective dynamics of 'small-world' networks.
\newblock {\em Nature}, 393:440--442, 1998.

\bibitem{young07dotprod} S.~J. Young and E.~R. Scheinerman.
\newblock {\em {R}andom {D}ot {P}roduct {G}raph {M}odels for {S}ocial {N}etworks}, volume 4863 of {\em Lecture Notes in Computer Science}.
\newblock 2007.

\end{thebibliography}

\appendix
\section{Variational EM Algorithm}

In Section~\ref{sec:problem}, we proposed a version of \MAG~by introducing a generative Bernoulli model for node attributes and formulated the problem to solve.
In the following Section~\ref{sec:algorithm}, we gave a sketch of \MAGFIT~that used the variational EM algorithm to solve the problem.
Here we provide how to compute the gradients of the model parameters ($\phi$, $\mu$, and $\Theta$) for the of E-step and M-step that we omitted in Section~\ref{sec:algorithm}.
We also give the details of the fast \MAGFIT~in the following.

\subsection{Variational E-Step}

In the E-step, the \MAG~parameters $\mu$ and $\Theta$ are given and we aim to find the optimal variational parameter $\phi$
that maximizes $\LQ(\mu, \Theta)$ as well as minimizes the mutual information factor $\mbox{MI}(F)$.
%
We randomly select a batch of entries in $\phi$ and update the selected entries by their gradient values of the objective function $\LQ(\mu, \Theta)$.
We repeat this updating procedure until $\phi$ converges.

In order to obtain $\nabla_{\phi} \left( \LQ(\mu, \Theta) - \lambda \MI(F) \right)$,
we compute $\PLQ{\phi_{il}}$ and $\frac{\partial \MI}{\partial \phi_{il}}$ in turn as follows.

\xhdr{Computation of $\PLQ{\phi_{il}}$}
To calculate the partial derivative $\PLQ{\phi_{il}}$, we begin by restating $\LQ(\mu, \Theta)$ as a function of one specific parameter $\phi_{il}$ and differentiate this function over $\phi_{il}$.
For convenience, we denote $F_{-il} = \{F_{jk}: j \neq i, k \neq l\}$
and
$
Q_{-il} = \prod_{j \neq i, k \neq l} Q_{jk}
$.
Note that $\sum_{F_{il}} Q_{il}(F_{il}) = 1$ and $\sum_{F_{-il}} Q_{-il}(F_{-il})$ because both are the sums of probabilities of all possible events.
Therefore, we can separate $\LQ(\mu, \Theta)$ in \EQ{eq:lqdef} into the terms of $Q_{il}(F_{il})$ and $Q_{-il}(F_{-il})$:
%
\begin{align}
& \LQ(\mu, \Theta) \nonumber \\
& = \EXATTR{}{\log P(A, F | \mu, \Theta) - \log Q(F)} \nonumber \\
& = \sum_{F} Q\left(F\right) \left( \log \LIKJOIN - \log Q\left(F\right) \right) \nonumber \\
& = \sum_{F_{-il}} \sum_{F_{il}} Q_{-il}(F_{-il}) Q_{il}(F_{il}) \nonumber \\
& \quad \times \left( \log \LIKJOIN - \log Q_{il}\left(F_{il}\right) - \log Q_{-il}\left(F_{-il}\right) \right) \nonumber \\
& = \sum_{F_{il}} Q_{il}(F_{il}) \left( \sum_{F_{-il}} Q_{-il}\left(F_{-il}\right) \log \LIKJOIN \right) \nonumber \\
& \quad \quad
- \sum_{F_{il}} Q_{il}(F_{il}) \log Q_{il}(F_{il})  \nonumber \\
& \quad \quad
- \sum_{F_{-il}} Q_{-il}(F_{-il}) \log Q_{-il}(F_{-il}) \nonumber \\
& = \sum_{F_{il}} Q_{il}(F_{il}) \EXATTR{-il}{\log \LIKJOIN} \nonumber \\
& \quad \quad + \mathcal{H} (Q_{il}) + \mathcal{H} (Q_{-il})
\label{eq:lqsep}
\end{align}
where $\mathcal{H}(P)$ represents the entropy of distribution $P$.
%

Since we compute the gradient of $\phi_{il}$, we regard the other variational parameter $\phi_{-il}$ as a constant so $\mathcal{H} (Q_{-il})$ is also a constant.
Moreover, as $\EXATTR{-il}{\log \LIKJOIN}$ integrates out all the terms
with regard to $\phi_{-il}$, 
it is a function of $F_{il}$.
Thus, for convenience, we denote $\EXATTR{-il}{\log \LIKJOIN}$ as $\log \PFUNC{F_{il}}$.
Then, since $F_{il}$ follows a Bernoulli distribution with parameter $\phi_{il}$, by \EQ{eq:lqsep}
\begin{align}
\LQ(\mu, \theta) & = (1 - \phi_{il}) \left( \log \PFUNC{1} - \log (1-\phi_{il}) \right) \nonumber \\
& \quad + \phi_{il} \left( \log \PFUNC{0} - \log \phi_{il} \right) + const \, .
\label{eq:lqonephi}
\end{align}

%
%
Note that both $\PFUNC{0}$ and $\PFUNC{1}$ are constant. Therefore,
\begin{equation}
\PLQ{\phi_{il}} = \log \frac{\PFUNC{0}}{\phi_{il}} - \log \frac{\PFUNC{1}}{1-\phi_{il}} \, .
\label{eq:eqonephi}
\end{equation}
\hide{
From \EQ{eq:eqonephi}, the optimal $\phi_{il}$ is as follows:
\begin{equation}
\phi_{il} = \frac{\PFUNC{0}}{\PFUNC{0} + \PFUNC{1}} \, .
\label{eq:updatephi}
\end{equation}
}

To complete the computation of $\PLQ{\phi_{il}}$,
now we focus on the value of $\PFUNC{F_{il}}$ for $F_{il} = 0, 1$.
By \EQ{eq:jointlik}~and the linearity of expectation,
$\log \PFUNC{F_{il}}$ is separable into small tractable terms as follows:
\begin{align}
\log \PFUNC{F_{il}}
& = \EXATTR{-il}{\log P(A, F | \mu, \Theta)} \nonumber \\
& = \sum_{u, v} \EXATTR{-il}{\log P(A_{uv} | F_{u}, F_{v}, \Theta)} \nonumber \\
& \quad + \sum_{u, k} \EXATTR{-il}{\log P(F_{uk} | \mu_{k})}
\label{eq:ptildesep}
\end{align}
where $F_{i} = \{F_{il}: l = 1, 2, \cdots, L \}$.
However, if $u, v \neq i$, then $\EXATTR{-il}{\log P(A_{uv} | F_{u}, F_{v}, \Theta)}$ is a constant,
because the average over $Q_{-il}(F_{-il})$ integrates out all the variables $F_{u}$ and $F_{v}$.
Similarly, if $u \neq i$ and $k \neq l$, then $\EXATTR{-il}{\log P(F_{uk} | \mu_{k})}$ is a constant.
Since most of terms in \EQ{eq:ptildesep}~are irrelevant to $\phi_{il}$, $\log \PFUNC{F_{il}}$~is simplified as
\begin{align}
\log \PFUNC{F_{il}}
& = \left(\sum_{j} \EXATTR{-il}{\log P(A_{ij} | F_{i}, F_{j}, \Theta)}\right) \nonumber \\
& \quad + \left(\sum_{j} \EXATTR{-il}{\log P(A_{ji} | F_{j}, F_{i}, \Theta)}\right)  \nonumber \\
& \quad + \log P(F_{il} | \mu_{l}) + C 
\label{eq:ptildesim}
\end{align}
for some constant $C$.

By definition of $P(F_{il} | \mu_{l})$ in \EQ{eq:jointlik}, the last term in \EQ{eq:ptildesim} is 
\begin{equation}
\log P(F_{il} | \mu_{l}) = F_{il} \log \mu_{l} + (1-F_{il}) \log (1 - \mu_{il})
\, .
\label{eq:logmu}
\end{equation}

With regard to the first two terms in \EQ{eq:ptildesim},
\[
\log P(A_{ij} | F_{i}, F_{j}, \Theta)
= \log P(A_{ji} | F_{i}, F_{j}, \Theta^{T}) \, .
\]
Hence, the methods to compute the two terms are equivalent.
Thus, we now focus on the computation of $\EXATTR{-il}{\log P(A_{ij} | F_{i}, F_{j}, \Theta)}$.

First, in case of $A_{ij} = 1$, by definition of $P(A_{ij} | F_{i}, F_{j})$ in \EQ{eq:jointlik},
\begin{align}
& \EXATTR{-il}{\log P(A_{ij} = 1 | F_{i}, F_{j}, \Theta)}  \nonumber \\
& = \EXATTR{-il}{\sum_{k} \log \Theta_{k} [F_{ik}, F_{jk}]} \nonumber \\
& = \EXATTR{jl}{\log \Theta_{l}[F_{il}, F_{jl}]} + \sum_{k \neq l} \EXATTR{ik, jk}{\log \Theta_{k} [F_{ik}, F_{jk}]} \nonumber \\
& = \EXATTR{jl}{\log \Theta_{l}[F_{il}, F_{jl}]} + C'
\label{eq:logedgeprob1}
\end{align}
for some constant $C'$ where $Q_{ik, jk}(F_{ik}, F_{jk}) = Q_{ik}(F_{ik}) Q_{jk}(F_{jk})$,
because $\EXATTR{ik, jk}{\log \Theta_{k}[F_{ik}, F_{jk}]}$ is constant for each $k$.

Second, in case of $A_{ij} = 0$,
\begin{align}
P(A_{ij} = 0 | F_{i}, F_{j}, \Theta)
& = 1 - \prod_{k} \Theta_{k} [F_{ik}, F_{jk}]
\, .
\label{eq:logedgeprob0}
\end{align}
Since $\log P(A_{ij} | F_{i}, F_{j}, \Theta)$ is not separable in terms of $\Theta_{k}$, it takes $O(2^{2L})$ time to compute $\EXATTR{-il}{\log P(A_{ij} | F_{i}, F_{j}, \Theta)}$ exactly.
We can reduce this computation time to $O(L)$ by applying Taylor's expansion of $\log (1 - x) \approx -x - \frac{1}{2}x^{2}$ for small $x$:
\begin{align}
& \EXATTR{-il}{\log P(A_{ij} = 0 | F_{i}, F_{j}, \Theta)}  \nonumber \\
& \approx \EXATTR{-il}{-\prod_{k} \Theta_{k}[F_{ik}, F_{jk}] - \frac{1}{2} \prod_{k} \Theta^{2}_{k}[F_{ik}, F_{jk}]} \nonumber \\
& = -\EXATTR{jl}{\Theta_{l}[F_{il}, F_{jl}]} \prod_{k \neq l} \EXATTR{ik, jk}{\Theta_{k}[F_{ik}, F_{jk}]} \nonumber \\
& \quad \quad - \frac{1}{2} \EXATTR{jl}{\Theta^{2}_{l}[F_{il}, F_{jl}]} \prod_{k \neq l} \EXATTR{ik, jk}{\Theta^{2}_{k}[F_{ik}, F_{jk}]}
\label{eq:logedgeprob0approx}
\end{align}
where each term can be computed by
\begin{align*}
\EXATTR{il}{Y_{l}[F_{il}, F_{jl}]} & = \phi_{jl} Y_{l}[F_{il}, 0] + (1-\phi_{jl}) Y_{l}[F_{il}, 1] \\
\EXATTR{ik, jk}{Y_{k}[F_{ik}, F_{jk}]} & = [\phi_{ik}~~\phi_{jk}] \cdot Y_{k} \cdot [1-\phi_{ik} ~~ 1-\phi_{jk}]^{T}
\end{align*}
for any matrix $Y_{l}, Y_{k} \in \mathbb{R}^{2 \times 2}$.

In brief, for fixed $i$ and $l$, we first compute $\EXATTR{-il}{\log P(A_{ij} | F_{i}, F_{j}, \Theta)}$ for each node $j$ depending on whether or not $i \rightarrow j$ is an edge.
By adding $\log P(F_{il} | \mu_{l})$, we then acheive the value of $\log \PFUNC{F_{il}}$ for each $F_{il}$.
Once we have $\log \PFUNC{F_{il}}$, we can finally compute $\PLQ{il}$.

\xhdr{Scalable computation}
However, as we analyzed in Section~\ref{sec:algorithm}, the above E-step algorithm requires $O(LN)$ time for each computation of $\PLQ{\phi_{il}}$ so that the total computation time is $O(L^{2}N^{2})$, 
which is infeasible when the number of nodes $N$ is large.

Here we propose the scalable algorithm of computing $\PLQ{\phi_{il}}$
by further approximation.
As described in Section~\ref{sec:algorithm}, we quickly approximate the value of $\PLQ{\phi_{il}}$ as if the network would be empty, 
and adjust it by the part where edges actually exist.
To approximate $\PLQ{\phi_{il}}$ in empty network case, 
we reformulate the first term in \EQ{eq:ptildesim}:
\begin{align}
&  \sum_{j} \EXATTR{-il}{\log P(A_{ij} | F_{i}, F_{j}, \Theta)} = \sum_{j} \EXATTR{-il}{\log P(0 | F_{i}, F_{j}, \Theta)} \nonumber \\
& \quad + \sum_{A_{ij} = 1} \EXATTR{-il}{\log P(1 | F_{i}, F_{j}, \Theta) - \log P(0 | F_{i}, F_{j}, \Theta)}
\label{eq:fastestep1_x}
\end{align}
However, since the sum of \textit{i.i.d.} random variables can be approximated in terms of the expectaion of the random variable,
the first term in \EQ{eq:fastestep1_x} can be approximated as follows:
\begin{align}
& \sum_{j} \EXATTR{-il}{\log P(0 | F_{i}, F_{j}, \Theta)}  \nonumber \\
& = \EXATTR{-il}{\sum_{j} \log P(0 | F_{i}, F_{j}, \Theta)} \nonumber \\
& \approx \EXATTR{-il}{(N-1) \mathbb{E}_{F_{j}}[ \log P(0 | F_{i}, F_{j}, \Theta)]} \nonumber \\
& = (N-1) \mathbb{E}_{F_{j}} [\log P(0 | F_{i}, F_{j}, \Theta)]
\label{eq:fastestep2_x}
\end{align}
As $F_{jl}$ marginally follows a Bernoulli distribution with $\mu_{l}$, we can compute \EQ{eq:fastestep2_x} by using \EQ{eq:logedgeprob0approx} in $O(L)$ time.
Since the second term of \EQ{eq:fastestep1_x} takes $O(L N_{i})$ time where $N_{i}$ represents the number of neighbors of node $i$,
\EQ{eq:fastestep1_x} takes only $O(L N_{i})$ time in total.
As in the E-step we do this operation by iterating for all $i$'s and $l$'s,
the total computation time of the E-step eventually becomes $O(L^2 E)$,
which is feasible in many large-scale networks.

\xhdr{Computation of $\PMI{\phi_{il}}$}
Now we turn our attention to the derivative of the mutual information term.
Since $\MI(F) = \sum_{l \neq l'} \MI_{ll'}$,
we can separately compute the derivative of each term $\frac{\partial \MI_{ll'}}{\partial \phi_{il}}$.
By definition in \EQ{eq:midef} and Chain Rule,
\begin{align}
& \frac{\partial \MI_{ll'}}{\partial \phi_{il}} = 
\sum_{x, y \in \{0, 1\}} \frac{\partial p_{ll'}(x,y)}{\partial \phi_{il}} \log \frac{p_{ll'}(x, y) }{p_{l}(x)p_{l'}(y)} \nonumber \\
& \quad + \frac{\partial p_{ll'}(x, y)}{\partial \phi_{il}} 
+ \frac{p_{ll'}(x, y)}{p_{l}(x)} \frac{\partial p_{l}(x)}{\partial \phi_{il}} 
+ \frac{p_{ll'}(x, y)}{p_{l'}(y)} \frac{\partial p_{l'}(y)}{\partial \phi_{il}} \,.
\label{eq:partialmi}
\end{align}
The values of $p_{ll'}(x, y)$, $p_{l}(x)$, and $p_{l'}(y)$ are defined in \EQ{eq:midef}.
Therefore, in order to compute $\frac{\partial \MI_{ll'}}{\partial \phi_{il}}$,
we need the values of $\frac{\partial p_{ll'}(x, y)}{\partial \phi_{il}}$, $\frac{\partial p_{l}(x)}{\partial \phi_{il}}$, and $\frac{\partial p_{l'}(y)}{\partial \phi_{il}}$.
By definition in \EQ{eq:midef},
\begin{align*}
& \frac{\partial p_{ll'}(x, y)}{\partial \phi_{il}} = Q_{il'}(y) \frac{\partial Q_{il}}{\partial \phi_{il}} \\
& \frac{\partial p_{l}(x)}{\partial \phi_{il}} = \frac{\partial Q_{il}}{\partial \phi_{il}} \\
& \frac{\partial p_{l'}(y)}{\partial \phi_{il}} = 0
\end{align*}
where $\frac{\partial Q_{il}}{\partial \phi_{il}} |_{F_{il}=0} = 1$
and $\frac{\partial Q_{il}}{\partial \phi_{il}} |_{F_{il}=1} = -1$.

Since all terms in $\frac{\partial \MI_{ll'}}{\partial \phi_{il}}$ are tractable, we can eventually compute $\PMI{\phi_{il}}$.




\subsection{Variational M-Step}

In the E-Step, 
with given model parameters $\mu$ and $\Theta$, we updated the variational parameter $\phi$
to maximize $\LQ(\mu, \Theta)$ as well as to minimize the mutual information between every pair of attributes.
In the M-step,
we basically fix the approximate posterior distribution $Q(F)$, \ie~fix the variational parameter $\phi$,
and update the model parameters $\mu$ and $\Theta$ to maximize $\LQ(\mu, \Theta)$.

To reformulate $\LQ(\mu, \Theta)$
by \EQ{eq:jointlik},
\begin{align}
& \LQ(\mu, \Theta) \nonumber \\
& = \EXATTR{}{\log P(A, F | \mu, \Theta) - \log Q(F)} \nonumber \\
& = \EXATTR{}{\sum_{i, j} P(A_{ij} | F_{i}, F_{j}, \Theta)
+ \sum_{i, l} P(F_{il} | \mu_{l}) }  + \mathcal{H}(Q) \nonumber \\
& = \sum_{i, j} \EXATTR{i, j}{\log P(A_{ij} | F_{i}, F_{j}, \Theta)} \nonumber \\
& \quad + \sum_{l} \left( \sum_{i} \EXATTR{il}{\log P(F_{il} | \mu_{l})} \right)
+ \mathcal{H}(Q)
\label{eq:mstep}
\end{align}
where $Q_{i, j}(F_{\{i\cdot\}}, F_{\{j\cdot\}})$ represents $\prod_{l} Q_{il}(F_{il}) Q_{jl}(F_{jl})$.

After all, $\LQ(\mu, \Theta)$ in \EQ{eq:mstep} is divided into the following terms:
%
a function of $\Theta$, a function of $\mu_{l}$, and a constant.
Thus, we can exclusively update $\mu$ and $\Theta$. Since we already showed how to update $\mu$ in Section~\ref{sec:algorithm}, here we focus on the maximization of $\LTH = \EXATTR{}{\log P(A, F | \mu, \Theta) - \log Q(F)}$ using the gradient method.


\xhdr{Computation of $\nabla_{\Theta_{l}} \LTH$}
To use the gradient method, we need to compute the gradient of $\LTH$:
\begin{align}
\nabla_{\Theta_{l}} \LTH = \sum_{i, j} \nabla_{\Theta_{l}} \EXATTR{i, j}{\log P(A_{ij} |F_{i}, F_{j}, \Theta)} \, .
\label{eq:mstep-theta_x}
\end{align}
We separately calculate the gradient of each term in $\LTH$ as follows:
%
For every $z_{1}, z_{2} \in \{0, 1\}$,
%
if $A_{ij} = 1$,
\begin{align}
& \frac{\partial \EXATTR{i, j}{\log P(A_{ij} | F_{i}, F_{j}, \Theta)}}{\partial \Theta_{l}[z_{1}, z_{2}]} \bigg|_{A_{ij} = 1} \nonumber \\
& = \frac{\partial}{\partial \Theta_{l}[z_{1}, z_{2}]}
\EXATTR{i, j}{\sum_{k} \log \Theta_{k}[F_{ik}, F_{jk}]} \nonumber \\
& = \frac{\partial}{\partial \Theta_{l}[z_{1}, z_{2}]} \EXATTR{i, j}{\log \Theta_{l}[F_{il}, F_{jl}]} \nonumber \\
& = \frac{Q_{il}(z_{1}) Q_{jl}(z_{2})}{\Theta_{l}[z_{1}, z_{2}]} \, .
\label{eq:mstep-theta1}
\end{align}
%
On the contrary, if $A_{ij} = 0$,
we use Taylor's expansion as used in \EQ{eq:logedgeprob0approx}:
\begin{align}
& \frac{\partial \EXATTR{i, j}{\log P(A_{ij} | F_{i}, F_{j}, \Theta)}}{\partial \Theta_{l}[z_{1}, z_{2}]} \bigg|_{A_{ij} = 0} \nonumber \\
& \approx \frac{\partial}{\partial \Theta_{l}} \EXATTR{i, j}{- \prod_{k} \Theta_{k}[F_{ik}, F_{jk}] - \frac{1}{2} \prod_{k} \Theta^{2}_{k}[F_{ik}, F_{jk}]} \nonumber \\
& = -Q_{il}(z_{1})Q_{jl}(z_{2}) \prod_{k \neq l} \EXATTR{ik, jk}{\Theta_{k}[F_{ik}, F_{jk}]} \nonumber \\
& \quad \quad  - Q_{il}(z_{1})Q_{jl}(z_{2})\Theta_{k}[z_{1}, z_{2}] \prod_{k \neq l} \EXATTR{ik, jk}{\Theta^{2}_{k}[F_{ik}, F_{jk}]}
\label{eq:mstep-theta2}
\end{align}
where $Q_{il, jl}(F_{il}, F_{jl}) = Q_{il}(F_{il}) Q_{jl}(F_{jl})$.

Since
\[
\EXATTR{ik, jk}{f(\Theta)} = \sum_{z_{1}, z_{2}} Q_{ik}(z_{1})Q_{jk}(z_{2})f\left(\Theta[z_{1}, z_{2}]\right)
\]
for any function $f$
and we know each function values of $Q_{il}(F_{il})$ in terms of $\phi_{il}$,
we are able to achieve the gradient $\nabla_{\Theta_{l}}\LTH$ by \EQ{eq:mstep-theta_x}~$\sim$~(\ref{eq:mstep-theta2}).

\xhdr{Scalable computation}
%
The M-step requires to sum $O(N^{2})$ terms in \EQ{eq:mstep-theta_x} where each term takes $O(L)$ time to compute.
Similarly to the E-step, here we propose the scalable algorithm by separating \EQ{eq:mstep-theta_x} into two parts, the fixed part for an empty graph and the adjustment part for the actual edges:
\begin{align}
& \nabla_{\Theta_{l}} \LTH =
\sum_{i, j} \nabla_{\Theta_{l}} \EXATTR{i, j}{\log P(0 |F_{i}, F_{j}, \Theta)} \nonumber \\
& \quad + \sum_{A_{ij} = 1} \nabla_{\Theta_{l}} \EXATTR{i, j}{\log P(1 |F_{i}, F_{j}, \Theta) - \log P(0 |F_{i}, F_{j}, \Theta)} \, .
\label{eq:fastmstep1_x}
\end{align}
We are able to approximate the first term in \EQ{eq:fastmstep1_x}, the value for the empty graph part, as follows:
\begin{align}
& \sum_{i, j} \nabla_{\Theta_{l}} \EXATTR{i, j}{\log P(0 | F_{i}, F_{j}, \Theta)} \nonumber \\
& = \nabla_{\Theta_{l}} \EXATTR{i, j}{\sum_{i, j} \log P(0 | F_{i}, F_{j}, \Theta)} \nonumber \\
& \approx \nabla_{\Theta_{l}} \EXATTR{i, j}{N(N-1) \mathbb{E}_{F} [\log P(0 | F, \Theta)]} \nonumber \\
& = \nabla_{\Theta_{l}} N(N-1) \mathbb{E}_{F} [\log P(0 | F, \Theta)] \,.
\label{eq:fastmstep2_x}
\end{align}
Since each $F_{il}$ marginally follows the Bernoulli distribution with $\mu_{l}$, \EQ{eq:fastmstep2_x} is computed by \EQ{eq:mstep-theta2} in $O(L)$ time.
As the second term in \EQ{eq:fastmstep1_x} requires only $O(LE)$ time,
the computation time of the M-step is finally reduced to $O(LE)$ time.


\section{Experiments}

\subsection{Yahoo!-Ansers Network}

\begin{figure}[t]
  \centering
  \subfigure[In-degree]{\includegraphics[width=0.235\textwidth]{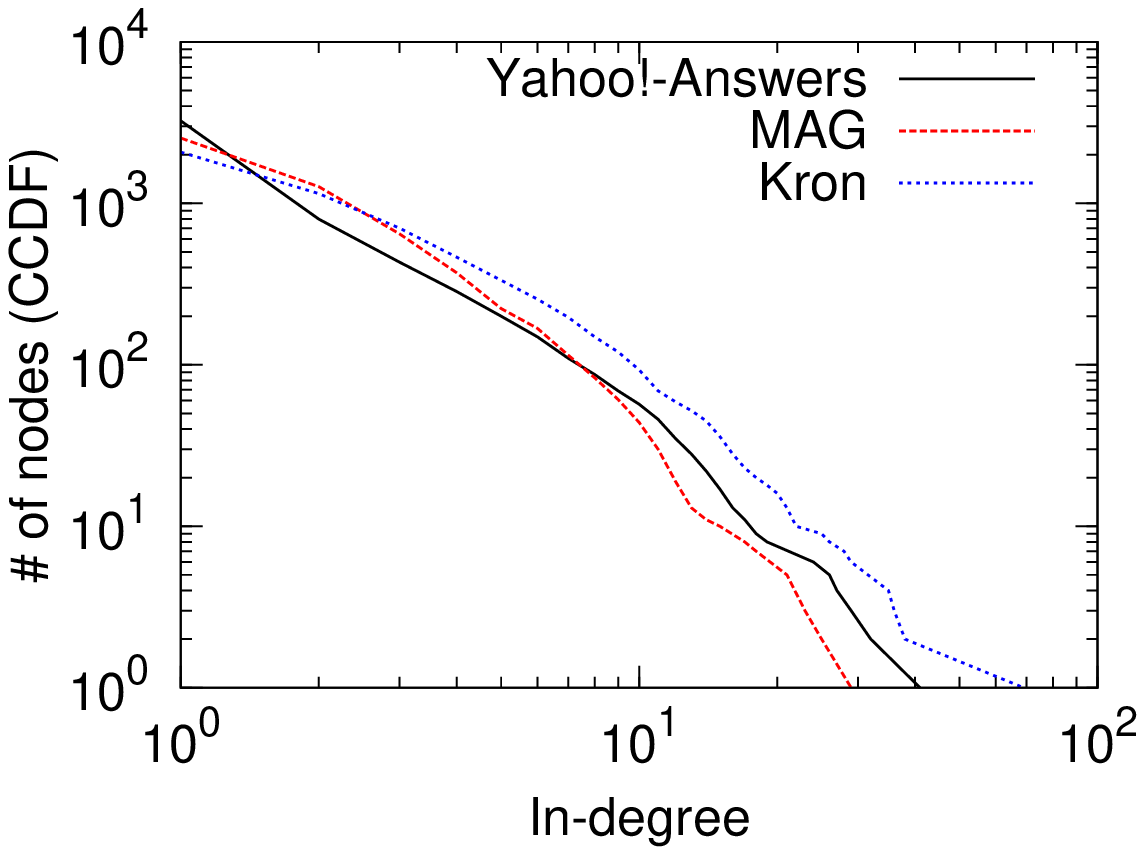}}
  \subfigure[Out-degree]{\includegraphics[width=0.235\textwidth]{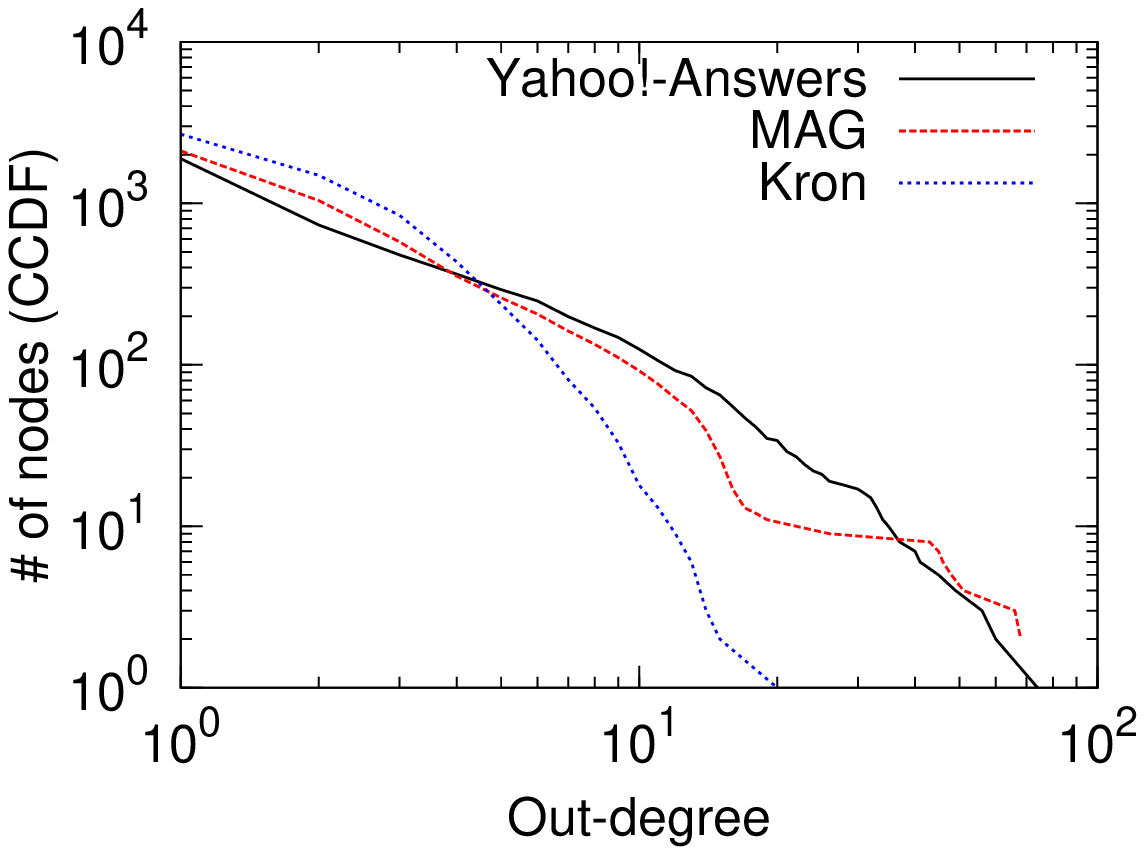}}
  \subfigure[Singular value]{\includegraphics[width=0.235\textwidth]{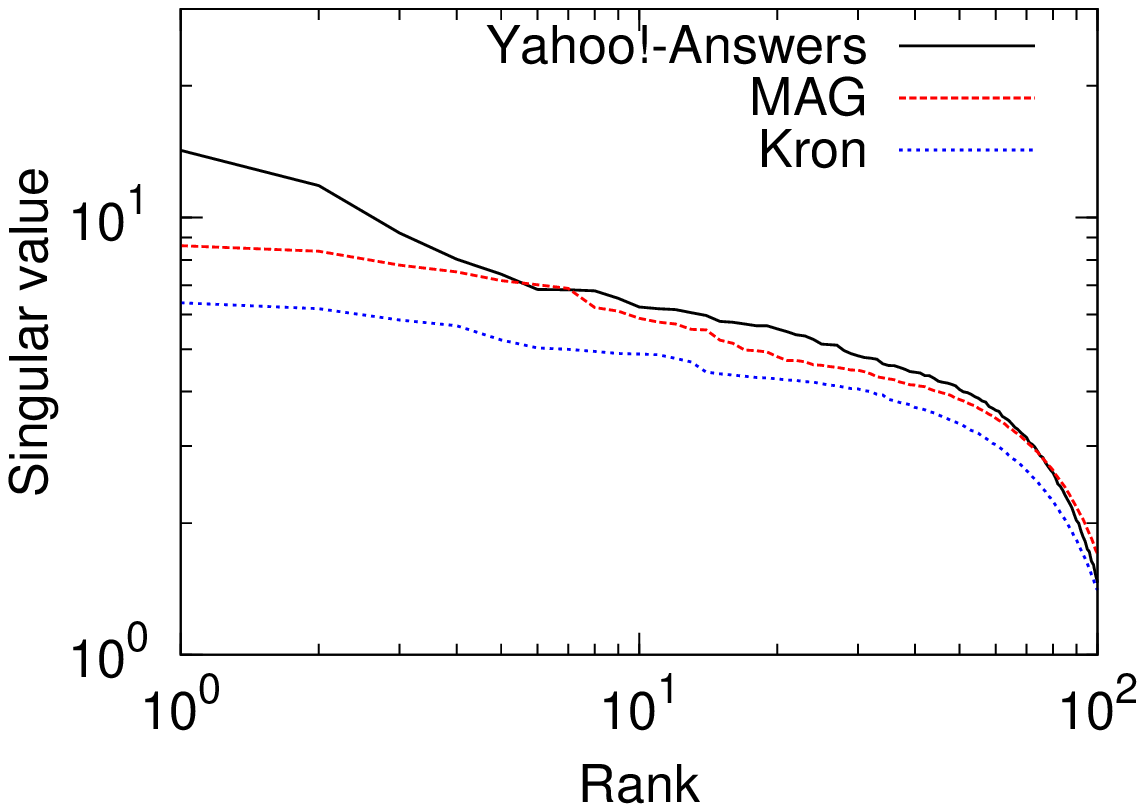}}
  \subfigure[Singular vector]{\includegraphics[width=0.235\textwidth]{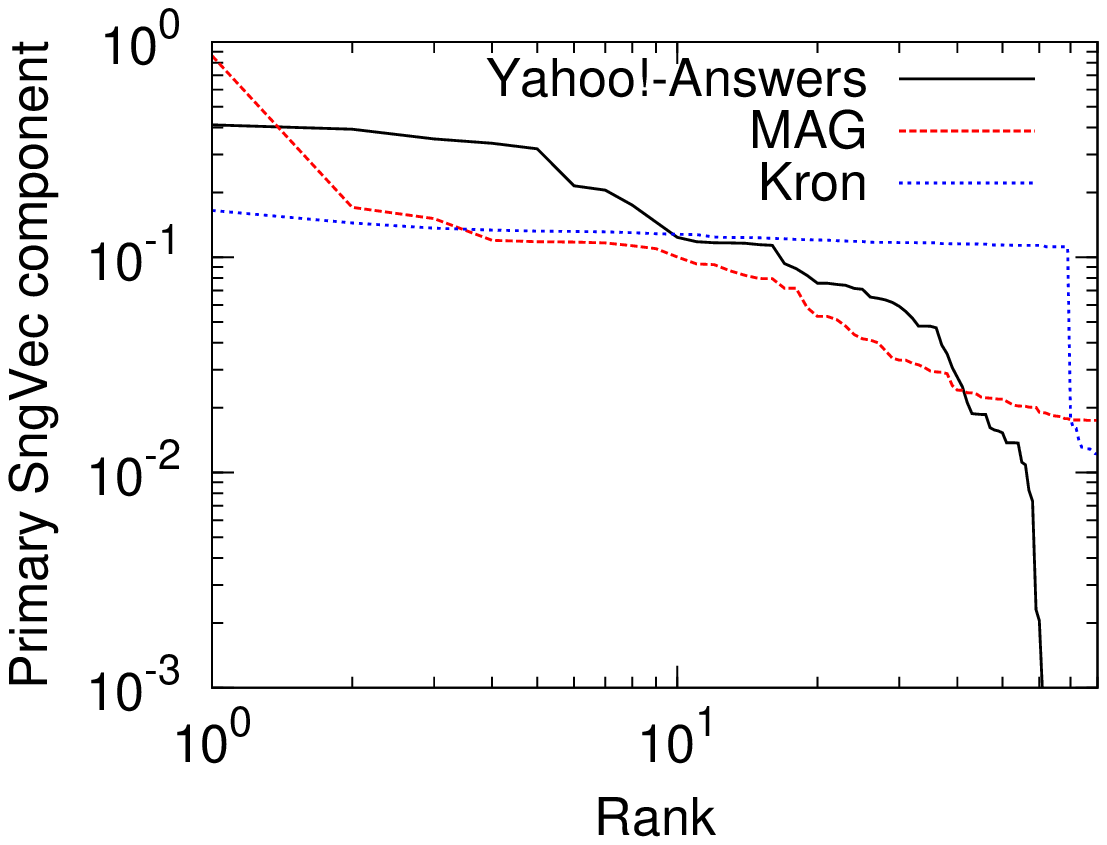}}
  \subfigure[Clustering coefficient]{\includegraphics[width=0.235\textwidth]{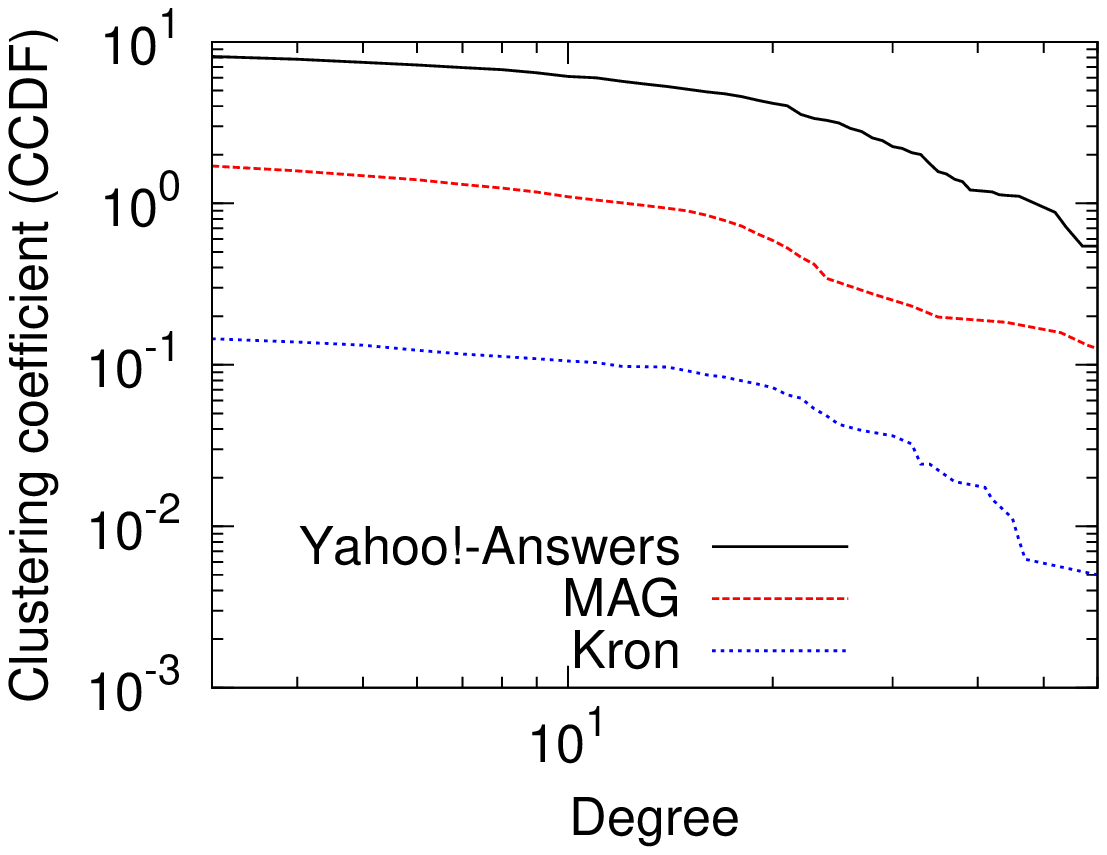}}
  \subfigure[Triad participation]{\includegraphics[width=0.235\textwidth]{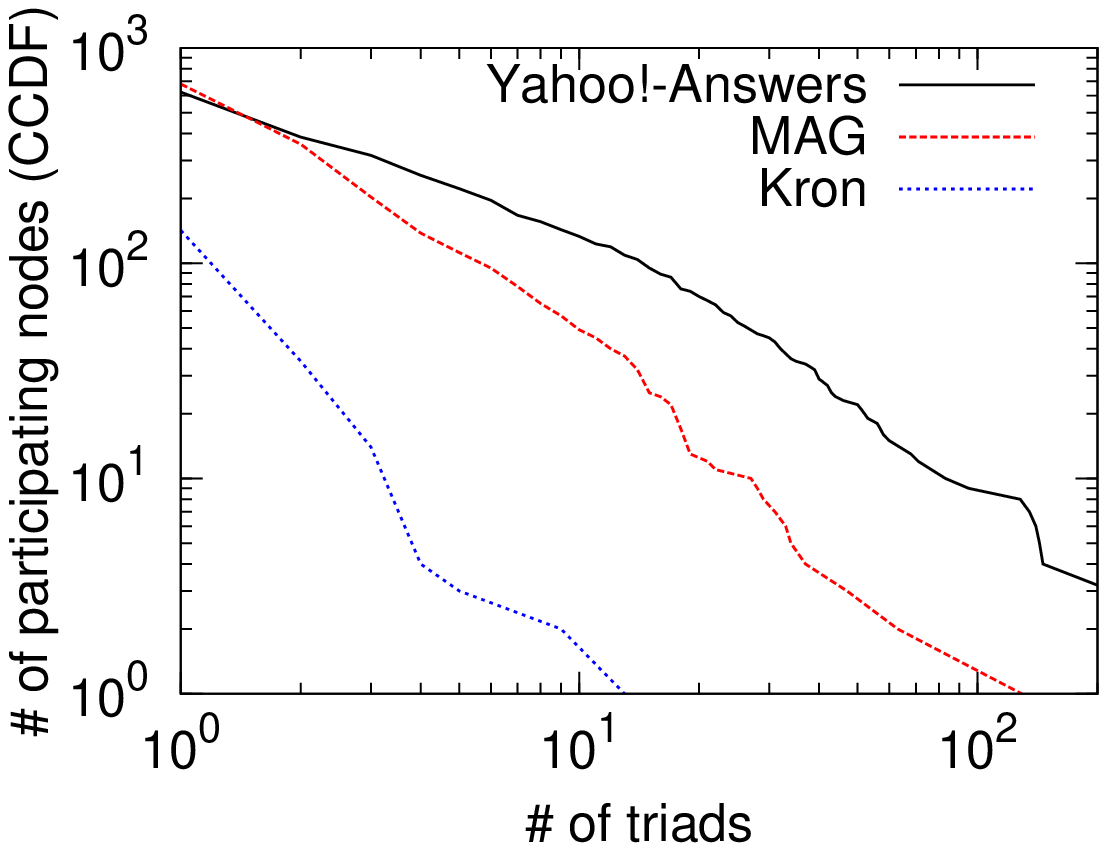}}
  \caption{The recovered network properties by the \MAG~and the Kronecker graphs
  model on the Yahoo!-Answers network.
  For every network property, \MAG~outperforms the Kronecker graphs model.}
  \label{fig:answerplot}
\end{figure}

Here we add some experimental results that we omitted in Section~\ref{sec:experiments}.
First, Figure~\ref{fig:answerplot} compares the six network properties of Yahoo!-Answers network and the synthetic networks generated by \MAG~and Kronecker graphs model fitted to the real network.
The \MAG~in general shows better performance than the Kronecker graphs model. Particularly, the \MAG~greatly outperforms the Kronecker graphs model in local-clustering properties (clustering coefficient and triad participation).

Second, to quantify the recovery of the network properties, we show the \KS~and \POWL~statistics for the synthetic networks generated by \MAG~and Kronecker graphs model in Table~\ref{tbl:answerks}.
Through Table~\ref{tbl:answerks}, we can confirm the visual inspection in Figure~\ref{fig:answerplot}.
The \MAG~shows better statistics than the Kronecker graphs model in overall
and there is huge improvement in the local-clustering properties.

\begin{table}
\caption{\KS~and \POWL~for MAG and Kronecker model fitted to Yahoo!-Answers network}
\label{tbl:answerks}
\centering \small
\begin{tabular}{c||c|c|c|c|c|c||c}
\multicolumn{8}{l}{\small{}}\\
  {\bf \KS} & InD & OutD & SVal & SVec & TP & CCF & Avg \\ \hline \hline
MAG  & 3.00 & 2.80 & 14.93 & 13.72 & 4.84 & 4.80 & 7.35 \\ \hline
Kron & 2.00 & 5.78 & 13.56 & 15.47 & 7.98 & 7.05 & 8.64 \\ \hline
  \multicolumn{8}{l}{\bf \POWL} \\ \hline \hline
MAG  & 0.96 & 0.74 & 0.70 & 6.81 & 2.76 & 2.39 & 2.39 \\ \hline
Kron & 0.81 & 2.24 & 0.69 & 7.41 & 6.14 & 4.73 & 3.67 \\ \hline
\end{tabular}
\end{table}

\subsection{AddHealth Network}

We briefly mentioned the logistic regression method in AddHealth network experiment.
Here we provide the details of the logistic regression and full experimental results of it.

For the variables of the logistic regression, we use a set of real attributes in the AddHealth network dataset.
For such set of attributes, we used F7 (forward selection) and R7 (random selection) defined in Section~\ref{sec:experiments}.
Once the set of attributes is fixed, we come up with a linear model:
\begin{align*}
P(i \rightarrow j) = 
\frac{\exp(c + \sum_{l} \alpha_{l} F_{il} + \sum_{l} \beta_{l} F_{jl}) }
{1 + \exp(c + \sum_{l} \alpha_{l} F_{il} + \sum_{l} \beta_{l} F_{jl}) } \,.
\end{align*}

Table~\ref{tbl:addhealthlogit} shows the \KS~and \POWL~statistics for logistic regression methods under R7 and F7 attribute sets.
It seems that the logistic regression succeeds in the recovery of degree distributions.
However, it fails to recover the local-clustering properties (clustering coefficient and triad participation) for both sets.

\begin{table}
\caption{\KS~and \POWL~for logistic regression methods fitted to AddHealth network}
\label{tbl:addhealthlogit}
\centering \small
\begin{tabular}{c||c|c|c|c|c|c||c}
\multicolumn{8}{l}{\small{}}\\
  {\bf \KS} & InD & OutD & SVal & SVec & TP & CCF & Avg \\ \hline \hline
R7 & 2.00 & 2.58 & 0.58 & 3.03 & 5.39 & 5.91 & 3.24 \\ \hline
F7  & 1.59 & 1.59 & 0.52 & 3.03 & 5.43 & 5.91 & 3.00 \\ \hline
  \multicolumn{8}{l}{\bf \POWL} \\ \hline \hline
R7 & 0.54 & 0.58 & 0.29 & 1.09 & 3.43 & 2.42 & 1.39 \\ \hline
F7  & 0.42 & 0.24 & 0.27 & 1.12 & 3.55 & 2.09 & 1.28 \\ \hline
\end{tabular}
\end{table}

\end{document}